\documentclass[nonacm,screen, acmlarge,natbib=false]{acmart}
\usepackage[style=ACM-Reference-Format,backend=bibtex,sorting=none, style=numeric-comp, maxnames=50]{biblatex}
\AtBeginDocument{%
  \providecommand\BibTeX{{%
    \normalfont B\kern-0.5em{\scshape i\kern-0.25em b}\kern-0.8em\TeX}}}

\acmJournal{CSUR}

\usepackage{tabularx}
\usepackage{amsmath}
\usepackage{amsthm}
\usepackage{longtable}
\usepackage{float}
\usepackage{svg}

\usepackage{caption}
\usepackage{subcaption}

\AtBeginBibliography{\small}

\usepackage{xcolor}
\usepackage{enumitem}

\usepackage{multirow}
\usepackage{makecell}

\usepackage[frozencache,cachedir=.]{minted}
\newenvironment{code}{\captionsetup{type=listing}}{}

\begin{document}

\title{A Survey on RISC-V Security: Hardware and Architecture}
\author{Tao Lu}
\authornote{This research is fully self-sponsored by the author. It does not represent the views of Marvell Semiconductor Ltd.}
\email{taolyu6@gmail.com}
\affiliation{%
  \institution{Marvell Semiconductor Ltd.}
  \country{USA}
}

\renewcommand{\shortauthors}{T. Lu}

\begin{abstract}
The Internet of Things (IoT) is an ongoing technological revolution. Embedded processors are the processing engines of smart IoT devices. For decades, these processors were mainly based on the Arm instruction set architecture (ISA). In recent years, the free and open RISC-V ISA standard has attracted the attention of industry and academia and is becoming the mainstream. Many companies have already owned or are designing RISC-V processors. Many important operating systems and major tool chains have supported RISC-V. Data security and user privacy protection are common challenges faced by all IoT devices. In order to deal with foreseeable security threats, the RISC-V community is studying security solutions aimed at achieving a root of trust (RoT) and ensuring that sensitive information on RISC-V devices is not tampered with or leaked. 
Many RISC-V security research projects are underway, but the academic community has not yet conducted a comprehensive survey of RISC-V security solutions. The latest technology and future development direction of RICS-V security research are still unclear. In order to fill this research gap, this paper presents an in-depth survey on RISC-V security technologies. This paper summarizes the representative security mechanisms of RISC-V hardware and architecture.
Specifically, we first briefly introduce the background and development status of RISC-V, and compare the RISC-V mechanisms with the most relevant Arm mechanisms, highlighting their similarities and differences.
Then, we investigate the security research of RISC-V around the theme of hardware and architecture security. Our survey covers hardware and physical access security, hardware-assisted security units, ISA security extensions, memory protection, cryptographic primitives, and side-channel attack protection. Based on our survey, we predict the future research and development directions of RISC-V security. We hope that our research can inspire RISC-V researchers and developers.
\end{abstract}



\maketitle
\section{Introduction}
The instruction set architecture (ISA) is an abstract model of a computer. ISA specifies the behavior of the running machine code without relying on specific machine implementations, thereby providing program compatibility between different implementations of the same architecture. The architecture specification conceptually defines the basic interface between hardware and software, the behaviors allowed for processor implementation, and the basic assumptions for software development and verification \cite{10.1145/3290384}.
Inspired by proprietary ISA's IP restrictions together with the lack of 64-bit addresses and overall complexity \cite{asanovic2014instruction}, RISC-V was developed and became more and more popular.
RISC-V aims to become a standard and universal ISA, especially for three representative device categories: small IoT devices, personal mobile devices, and warehouse-level computers.

After years of technological evolution, RISC-V has become a commercially available ISA. Currently, semiconductor companies are testing RISC-V, and even some product lines are transitioning to RISC-V. From the Google Scholar statistics of the RISC-V research work, we observe that the volume of RISC-V related research has exponentially increased in the past decade. The tremendous momentum of RISC-V adoption in computing platforms is very clear. Various RISC-V devices from small IoT microcontrollers to multi-core high-performance processors have been taped out. Recently, SiFive has taped out \emph{HiFive1 Rev B} and \emph{HiFive Unmatched} SoCs \cite{sifive_boards}, which have been put on market for IoT and desktop applications. SiFive RISC-V cores have already been used for SSD Controllers \cite{sifive_ssd_controller}. 
Alibaba has taped out \emph{Xuantie-910} for cloud and edge computing \cite{Chen2020Xuantie910}. Other tape-outs and FPGA boards include Microchip PolarFire SoC FPGA Icicle Kit, RISC-V multicore accelerator SoC BlackParrot \cite{petrisko2020blackparrot}, Xilinx multi-core FPGA system VC707 \cite{zhang2020parallel} etc. Lee et al. \cite{Lee2020mIoT} implemented a metamorphic IoT platform for on-demand hardware replacement in large-scaled IoT application scenarios.
All these post-silicon implementations push RISC-V from concepts to products.
European Processor Initiative (EPI), one of the cornerstones of the EuroHPC Joint Undertaking, a new European Union strategic entity focused on pooling the Union’s and national resources on HPC to build and deploy the most powerful supercomputers within Europe, is preparing to adopt RISC-V as its core solution for exascale embedded HPC platform \cite{Kovac2020}.

Table \ref{riscv_boards} lists existing post-silicon RISC-V chips, which are mainly used in smart devices and Internet of Things (IoT) devices. IoT consists of billions of connected devices, is changing fields such as medical care, transportation, and public services, and continues to collect, process, and transmit big data. IDC predicts that there will be 41.6 billion connected IoT devices by 2025, and the data generated will exceed 79ZB \cite{Microsoft2019}. The value of big data is widely recognized by the industry. The effective mining of big data can improve the competitive advantage of enterprises and provide a basis for the decision-making of social functional departments. However, the collection, storage, analysis and sharing of big data has brought new information security and privacy issues. Security has become a prominent challenge in the era of big data \cite{tankard2012big, kashyap2018impact, venkatraman2019big}. We expect that the security mechanism of the RISC-V architecture will play an important role in its future ecosystem.

With the development of cryptographic technology, many security mechanisms have been widely deployed in practice, including data confidentiality and integrity protection, identity verification, privacy protection, denial of service prevention, non-repudiation enforcement, and digital content protection.
Various security protocols and standards such as TLS, ZRTP, IPSec, IKE, and Kerberos have been used to protect data services and applications and alleviate platform security challenges. However, with technology advancement, the sophistication of attacks is developing simultaneously, especially cyber attacks are more complicated, illusory and more targeted than ever \cite{CPR2020}. In September 2019, iPhone hackers were exposed. For at least two years, attackers have used infected websites to Exploit 14 independent vulnerabilities in Apple iOS and install spyware on thousands of Apple devices that have visited websites infected with malware. Attackers can access regular user data, keychain passwords, and social media content \cite{apple2019}. Recently, FireEye, a publicly-listed cyber security company, was attacked by a highly sophisticated adversary who stole FireEye Red Team tools, which may be used for malicious cyber attacks on the system \cite{fireeye2020}. FireEye was forced to publish hundreds of countermeasures, So that the wider security community can protect itself from these tools. It is clear from the incident that no organization, whether a sophisticated security defender or not, is immune to destructive cyber attacks. According to a cyber security report released by Security Boulevard \cite{CB2020}, cybercrime caused approximately \$1.5 trillion in losses for victims in 2018.
AI-driven attacks \cite{yu2020ai, kaloudi2020ai} make the situation worse and make security defenses an increasingly serious challenge.
In addition to countless cyber attacks, the disclosures of Meltdown \cite{lipp2018meltdown} and Spectre \cite{kocher2019spectre} also revealed hardware vulnerabilities in modern processors. The Spectre and Meltdown attacks confirmed the need to treat security as a system-level design constraint that crosses the boundaries of hardware and software. The serious impact of ISA vulnerabilities has aroused unprecedented attention in the industry to architecture security. Side channel attacks have become widely known and have attracted a lot of research.

The market no longer only focuses on product performance, security is also a demanding requirement. 
At the application layer, security mechanisms for artificial intelligence, Internet of Things, and wireless sensor network platforms are proposed \cite{Williams-King2020, garofalo2019pulp, yang2020ai, bayerl2020offline, sisejkovic2020challenging, lee2020risc, banerjee2019energy, palmiero2018design, Takase2020A, gookyi2019selecting, Auer2019, duran2020energy}. At the system level, security mechanisms such as trusted boot \cite{kumar2020post,kumar2019itus,haj2019lightweight,8854418,9211555,both2020analyzing} and trusted execution \cite{dabbelt2018sifive, lee2020keystone, garlaticlean2020,de2015micro, nasahl2020hector, schrammel2020donky,andradesoftware2020,bahmani2020cure,kohlbrenner2020building,schneider2020pie} are widely used.
IoT systems usually run on small cores of embedded systems. These cores have low computing power and limited resources, making it difficult to implement complicated security policies. Due to the high efficiency of hardware execution, hardware-based security mechanisms can minimize the resource cost of these devices.
Arm TrustZone, Intel SGX, and AMD SEV technologies provide system-wide security solutions through hardware isolation implemented by the CPU \cite{pinto2019demystifying}. Although the existing hardware isolation technology is not impeccable, for example, the TEE system assisted by TrustZone has security vulnerabilities \cite{cerdeira2020sok}, it still plays an important role. Enhancing hardware and architecture security is an important requirement. Balancing platform security level and system performance, hardware and architecture security is important for platform security solutions. Therefore, chip suppliers are now actively introducing hardware security modules (HSM) and are very careful to avoid security vulnerabilities in hardware design.

%
 
The openness of RISC-V enables public auditing of the architecture design, thereby providing opportunities to build secure platforms. However, the openness of ISA provides attackers with more details behind the scenes, and system security vulnerabilities can be more easily discovered and exploited by adversaries. Therefore, RISC-V needs to use its openness to build reliable security mechanisms. The RISC-V security community needs to understand this relatively new architecture to carry out security technology innovations. RISC-V supports various privilege modes \cite{watermanrisc2019} and physical memory protection \cite{watermanrisc2019, cheangverifying2020, Kim2020}.
Trusted execution environments \cite{dabbelt2018sifive, lee2020keystone, garlaticlean2020,de2015micro, nasahl2020hector, schrammel2020donky,andradesoftware2020,bahmani2020cure,kohlbrenner2020building,schneider2020pie} have also been implemented. Other security enhancement measures including hardware security \cite{jin2015introduction, Song2015, Ma2019Dam, zang2019reconfigurable, menon2017shakti, banerjee2018energy, Liu2018, yu2020creating, batina2019hardware,gleissenthall2019iodine, de2020hardware, wistoff2020prevention, heiser2020towards, athalye2019notary}, memory protection \cite{malenko2019device, Kim2020, Savry2020,Song2015,menon2017shakti,Liu2018,Ma2019Dam,weiser2019timber,nienhuis2020rigorous}, ISA security extensions \cite{Li2020SIMF,escouteloup2020recommendations,yu2019data,wistoff2020prevention,heiser2020towards,watson2020capability,saarinen2019sneik,tehrani2019classification,Tehrani2020DSD,koppelmann2019risc,alkim2020isa,marshall2020design,zehrisc2020,Matsuoka2020Virtual}, cryptographic engines and primitives \cite{Nayak2017HOP,rao2018design,marshall2020design,saarinen2020building,Kiningham2019Falcon,Balagurusamy2019Crypto,Jager2020DICE,tehrani2019classification,loiseau2018binary,zehrisc2020,jellema2019optimizing,koppelmann2019risc, faruk2020embedded, steinegger2020fast, zhang2020transys, alkim2020isa, roy2020efficient, steinegger2020fast, Tehrani2020DSD, Campos2020, stoffelen2019efficient}, and side-channel prevention \cite{deng2019secure,gonzalez2019replicating,Napoli2018,Reinbrecht2020Guard,de2019protecting,zhang2018blacklist,Dao2020,van2020protecting,sauvage2017secure} have been proposed.
A systematic survey of the latest RISC-V security solutions will help the community understand the current state and future trends.
%

\begin{figure}[t]%
    \centering
    {\includesvg[width=13cm]{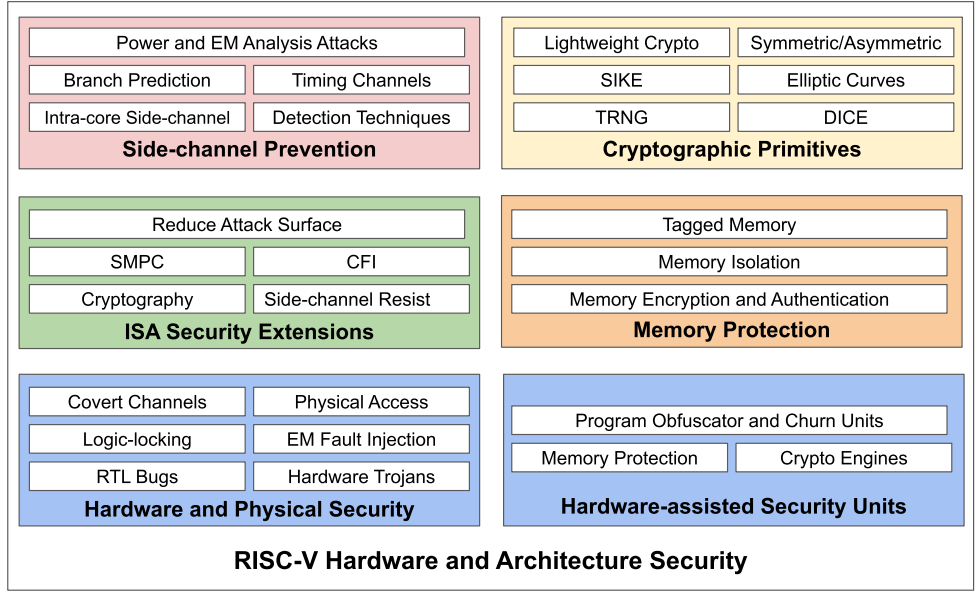}}
    \vspace{-2mm}
    \caption{The spectrum of RISC-V hardware and architecture security, which is the cornerstone of platform security.}%
\vspace{-4mm}
    \label{riscv_security_stack}%
\end{figure}

The organization of this article is as follows. First, we provide an overview of RISC-V security, focusing on discussing platform security requirements, the root of trust, and the building blocks of RISC-V architecture security, which lay a foundation for the enforcement of system and application security policies (\textbf{Section} \ref{riscv_overview}). Then, as it is summarized in Figure \ref{riscv_security_stack}, we categorize our survey into the following topics: hardware and physical security (\textbf{Section} \ref{subsect_hardware_physical_security}), hardware-assisted security units (\textbf{Section} \ref{hardware_assist_sec_unit}), memory protection (\textbf{Section} \ref{mem_protect}), ISA security extensions (\textbf{Section} \ref{ISA_Security_Extensions}), cryptographic primitives (\textbf{Section} \ref{crypto_primitive}), and protection against side-channel attacks (\textbf{Section} \ref{side_channel}).
Finally, we summarize our observations and discuss future directions of RISC-V security research (\textbf{Section} \ref{future_directions}). 

\begin{table}[t]
\renewcommand\arraystretch{1.0}
  \centering
  \caption{Representative RISC-V boards and their main features, application scenarios and operating system support.}
\vspace{-3mm}
\label{riscv_boards}
\addtolength{\tabcolsep}{-1.2pt}
\begin{tabular}{|>{\raggedright}p{2cm}|>{\raggedright}p{3cm}|>{\raggedright}p{1.6cm}|p{1.6cm}|p{1cm}|p{1.1cm}|p{3cm}|}
\hline
\textbf{Platforms} & \textbf{SoC \& Processor} & \textbf{Privilege Mode}&\textbf{Security \newline Feature}& \textbf{Arm Peer} &\textbf{Target App} &\textbf{Operating System Support}\\
\hline
\hline
\textbf{ICE EVB}  & XuanTie C910; 64-bit Dual cores 1.2GHz & U+S+M &PMP & Cortex-A55  & {5G, AI, Mobile} & Linux, Android \\
\hline
\textbf{HiFive Unmatched}  & SiFive U740; 64-bit Quad cores 1.4GHz & U+S+M &In-order; PMP & Cortex-A55  & {Generic PC} & Linux \\
\hline
\textbf{HiFive Unleashed}  & SiFive U54; 64-bit Quad cores 667MHz & U+S+M &In-order; PMP& Cortex-A53 &AI, IoT & Linux, \newline VxWorks \\
\hline
\textbf{BeagleV}  & SiFive U74; 64-bit Dual cores 1.0GHz & U+S+M & PMP & Cortex-A55 &AI & Linux, Zephyr \\
\hline
\textbf{PolarFire Icicle} & SiFive U54; 64-bit Quad cores 667MHz & U+S+M &In-order; PMP& Cortex-A53 &AI, IoT & Linux, seL4 \\
\hline
\textbf{Kendryte KD233 } & Kendryte K210; 64-bit Dual cores 400MHz & M & AES and SHA Accelerator & Cortex-M7& AI, IoT & FreeRTOS \\
\hline
\textbf{HiFive1 RevB} & SiFive E310; 32-bit core 320MHz & M & In-order; PMP & Cortex-M4 &IoT & Bare-metal, embOS, FreeRTOS, Mynewt, RT-Thread, Zephyr\\
\hline
\textbf{Gigadevice RV-STAR} & GD32VF103; 32-bit core 108MHz & M & NA & Cortex-M3 &Low-Power &Bare-metal \\
\hline
\end{tabular}
\vspace{0mm}
\end{table}
\section{An Overview of RISC-V Security}\label{riscv_overview}
In this chapter, we outline the security requirements and foundations of embedded platforms. We discuss the application and system security requirements (Section \ref{sec_requirements}), the hardware security module as the root of trust (Section \ref{RoT}), and the hardware and architectural foundations of the RISC-V security mechanism (Section \ref{building_blocks}).

\subsection{Security Requirements of General Platforms}\label{sec_requirements}
The goal of RISC-V is to become a general instruction set architecture. RISC-V has been used in low-power Internet of Things \cite{schiavone2017slow}, storage controllers \cite{sifive_ssd_controller,WD_2019_riscv}, artificial intelligence machine learning \cite{flamand2018gap, louis2019towards, garofalo2020pulp}, wireless sensor networks, data centers \cite{chen2020xuantie}, high-performance computing \cite{cavalcante2019ara}, and many other application scenarios. Table \ref{riscv_boards} lists the representative RISC-V boards on the market, and summarizes the processor models, security features, and target application scenarios. We can see from the table that most of the RISC-V SoCs are still used in low-power IoT devices. Recently, SiFive cooperated with Intel to release the Performance P550 core, which can be scaled up to quad-core complicated configurations, using an area similar to that of a single Arm Cortex-A75, while providing significant performance advantages per area \cite{sifive_p550}. We can expect that more and more high-performance RISC-V chips will be available in the future.
Many software and toolchains have been integrated into the RISC-V ecosystem. The RISC-V GNU Compiler Toolchain \cite{RISCV_GNU} and SiFive Freedom Studio IDE Toolchain \cite{sifive_freedom_studio2019} are two representative ones. Table \ref{riscv_boards} also lists operating systems that have supported the RISC-V boards. Linux and FreeRTOS are two important ones. In addition, many RISC-V SoCs can run bare-metal applications, which do not depend on an operating system.

RISC-V and Arm architecture compete with each other for similar application scenarios.
Mobile Internet is an important application scenario of Arm devices. Applications running on mobile devices such as smartphones and tablets increasingly rely on machine learning services to optimize user experience, such as estimating battery life based on user behavior, improving image quality, or performing voice recognition \cite{bayerl2020offline}. These services require frequent interaction with cloud servers, and the high sensitivity of such remotely processed data causes billions of users to face serious privacy risks. Recently, a British government contractor database containing more than 1 million fingerprints and facial recognition information was leaked \cite{2019face}, posing a major challenge to user privacy. Clients and service providers can use encryption technologies such as homomorphic encryption (HE) \cite{gentry2009fully} and secure multi-party computing (SMPC) \cite{goldreich1998secure} to securely process private inputs under encryption, or use provable security protocols to jointly compute any function on private inputs. Unfortunately, in the networking scenario the computational and network communication bottlenecks of performing complicated machine learning tasks greatly limit the usefulness of the above technologies. Processing all sensitive user data on-premise not only reduces the risk of data leakage, but also improves the performance of data processing. Therefore, exploring hardware-assisted solutions to provide secure and private complicated computing services directly on mobile devices is an important application requirement.

Due to the risk of tampering with system executable files via physical attacks, trusted boot is critical for system life cycle security. One of the principles of building a secure system is to generate a chain of trust from all software parts between the first bootloader to the last trusted application \cite{haj2019lightweight}. This chain of trust is based on a root of trust (RoT) that will never be easily tampered with. This is called the secure boot sequence. Many security devices including laptops, desktops, smart phones, and IoT devices, need to implement secure boot to ensure system integrity. The secure boot architecture is complicated, relying on code verification units to ensure the integrity of the chain of trust. The public key cryptography such as elliptic curve digital signature algorithm (ECDSA) and secure hash algorithm (SHA) are the basic primitives of secure boot, which are usually implemented in the RoT such as a hardware security module (HSM). We will discuss HSM and RoT in Section \ref{RoT}. 

Privilege management is the fundamental mechanism of architecture security. If the system can control tasks to run in privileged or non-privileged mode, and limit the task's access to resources such as RAM, executable code, and peripherals, it will make the microcontroller application more secure. For example, preventing certain code from executing in RAM can prevent attacks including buffer overflows and malicious code execution. However, implementing a memory protection mechanism \cite{bai2016arm} will make application design more complicated, because memory protection needs to determine the memory area limits and describe these limits to the operating system. Also, the memory protection mechanism requires differentiating operations and restrictions of applications. The memory protection strategy that restricts each task to its own memory area may be the safest, but the design and implementation are also the most complicated. Trusted Execution Environment (TEE) solutions for protecting sensitive code and data are widely deployed. Major CPU vendors have introduced their TEEs, such as Arm TrustZone \cite{pinto2019demystifying} and Intel SGX \cite{costan2016intel} to enable platform security zones. There are many application scenarios for TEE, including cloud servers, mobile phones, ISPs, IoT devices, sensors, and hardware tokens. TEE needs to be implemented based on hardware security building blocks including privilege management, memory protection, and even trusted boot \cite{weiser2019timber, lee2020keystone}. We will outline RISC-V security building blocks in the rest of this Section.
\begin{figure}[t]%
    \centering
    {\includesvg[width=10cm]{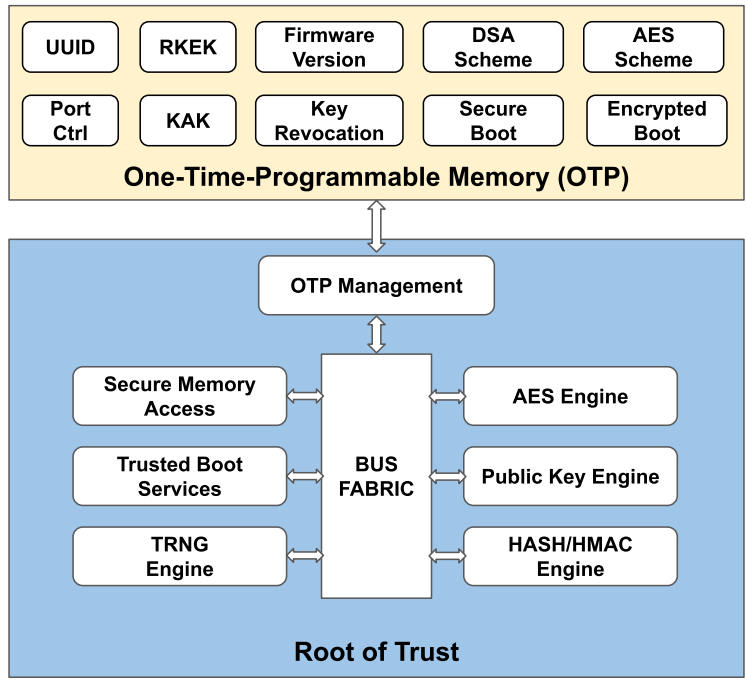}}
    \caption{The Building Blocks of the Root of Trust or Hardware Security Module.}%
\vspace{-4mm}
    \label{riscv_root_of_trust}%
\end{figure}
\subsection{The Root of Trust and the Hardware Security Module}\label{RoT}
The root of trust (RoT) \cite{eldefrawy2012smart} is the foundation on which all secure operations of computing systems depend. It contains keys for encryption functions, supports trusted boot and TEE. RoT is also important for public key infrastructure (PKI), which is used to generate and protect root and certificate authority keys, sign code for software security, immutability and authenticity, and create digital certificates for identity verification. Because the security of the system relies on the keys used to encrypt and decrypt data, as well as digital signatures and signature verification functions, the RoT is the always trusted source in the encryption system. RoT can secure data and applications and help build the chain of trust in the entire ecosystem. 

The RoT must be secure by design. The hardware-based RoT will not be attacked by malicious software, so it is the most secure. The RoT can be an independent security module or a security module in a system-on-chip (SoC). A fixed-function RoT is a state machine designed to perform a specific set of operations, such as data encryption, certificate verification, and key management. Usually these functions are static and can only perform their specially designed functions. In addition, there is a kind of programmable RoT. It is built around the CPU, can perform more complicated security functions, can be upgraded, and can run new encryption algorithms and security applications to counter evolving attack vectors. 

Since the RoT is the target of attackers, it is usually executed in isolation to ensure that sensitive security functions are executed in a dedicated security domain physically separated from the general-purpose processor. Safely isolating security functions in a physically separated RoT can reduce architecture complexity and optimize CPU performance. The RoT shall also have comprehensive anti-tampering and side-channel resistance capabilities, prevent fault injection and side-channel attacks, and support layered security to provide multiple layers of strong defense. For hardware-based roots of trust \cite{ehret2020hardware}, encryption engines, keys, and other sensitive security resources can only be accessed in hardware. Based on the hardware RoT, software security mechanisms can be implemented to provide additional flexibility. 

RoT solutions usually include a hardened hardware security module (HSM) that generates and protects keys and performs encryption functions in its secure environment \cite{wolf2011design}. HSM is a tamper-proof hardware device that can enhance system security. The HSM is usually used for platforms with high data security and trust, which is inaccessible outside the system, so the system can trust the authentic and authorized keys and other encrypted information received from the HSM. The HSM can pass various FIPS certifications to prove its security specifications. The implementation of the HSM and the RoT is complicated, involving hardware and architecture, including system permission level control, secure memory access, password instructions, random number generators, etc. As shown in Figure \ref{riscv_root_of_trust}, the RoT or HSM usually includes the following main components:
\begin{itemize}[leftmargin=*]
    \item \textbf{OTP Management Module} manages one-time programmable (OTP) memory \cite{robson2007electrically}, which is non-volatile and used to store keys and other security assets. OTP can be realized based on semiconductor anti-fuse and MOS gate oxide breakdown anti-fuse.
    OTP memory can only be programmed once, which is an irreversible process, thus ensuring security. The original equipment manufacturer (OEM) programs the OTP before the chip leaves the factory, and through it can write important trust-sensitive data, such as UUID, OEM key, firmware version, and trusted boot-related schemes, policies, and configuration parameters. OTP bears the RoT information in an immutable way to support the chain of trust throughout the chip life cycle.
    
    \item \textbf{Secure Memory} features multiple interfaces and a hardened memory protection unit. The RoT's RAM stores secure assets in a memory region isolated from the rest of the system. It may also include a small amount of ROM. Access to the memory regions is protected by the MPU\cite{arm-mpu-2016} or PMP \cite{cheangverifying_pmp} mechanisms to guarantee only entities with proper privilege levels can access the protected memory areas. We will discuss RISC-V secure memory related research in Section \ref{mem_protect}. 
    \item \textbf{Symmetric Cryptographic (eg. AES) Engine} is used for message and image file encryption and decryption to guarantee data confidentiality and support secure boot \cite{lebedev2018secure}. Symmetric encryption was the only type of encryption in use prior to the development of asymmetric cryptography in the 1970s. It remains by far more widely used because of higher performance and simpler key management. The Advanced Encryption Standard (AES) is a specification for the encryption of electronic data established by the U.S. National Institute of Standards and Technology (NIST) in 2001 \cite{nechvatal2001report}. Since then, AES has become the most widely used symmetric encryption algorithm. We will discuss RISC-V ISA extensions for AES in Section \ref{crypto_algorithm_ISE}.
    \item \textbf{Asymmetric Cryptographic \cite{diffie1988first} (eg. Public Key RSA) Engine} is used for message encryption and enables the sender to combine a message with a private key to create a short digital signature on the message. This scheme has the advantage of not having to share symmetric keys while gaining the higher data throughput advantage of symmetric-key cryptography.
    Asymmetric cryptographic systems use key pairs: a public key that may be known to others, and a private key that only the owner knows. The generation of this key pair depends on a cryptographic algorithm based on a mathematical one-way function. Effective security requires that the private key be kept private. The public key can be distributed. In such a system, anyone can use the public key of the target recipient to encrypt the message, but can only use the private key of the recipient to decrypt the encrypted message. Public key encryption can also perform reliable authentication by creating a short digital signature on the message. Public key algorithms are the basic security primitives in modern cryptographic systems, including applications and protocols that can guarantee the confidentiality, authenticity and non-repudiation of electronic communications and data storage. They are the basis of many Internet standards, such as the transport layer security protocol. Some public key algorithms provide key distribution and confidentiality (for example, Diffie-Hellman key exchange \cite{li2010research}), some provide digital signatures, and some provide both (for example, PKCS \cite{jonsson2003public}). Compared to symmetric encryption, asymmetric encryption is much slower than good symmetric encryption, which is too slow for many purposes. Today's cryptographic systems such as TLS use both symmetric encryption and asymmetric encryption. Asymmetric Cryptography is an essential building block for secure boot. 
    \item \textbf{HASH/HAMC Engine} is used for message hashing to guarantee data integrity and support secure boot. A hash function accepts a variable-length block of data as input and produces a fixed-size hash value, which can be used for message authentication, digital signatures, one-way password, and intrusion detection etc.  
    Hash-based message authentication code (HMAC) \cite{krawczyk1997hmac} is a specific type of message authentication code, involving cryptographic hash functions and secret keys. Like any MAC, it can be used to simultaneously verify data integrity and message authenticity. HASH and HMAC play an important role in various security applications and Internet protocols. In addition, the hash function is a necessary part of key derivation and public key algorithms such as PKCS \cite{jonsson2003public} and ECDSA \cite{johnson2001elliptic}.
    \item \textbf{Random Number/Bit Generation (eg. TRNG \cite{fischer2002true}) Engine} generates random numbers or bits for multiple cryptographic algorithms and protocols. Many network security algorithms and protocols based on cryptography use random values. For example, key distribution and mutual authentication schemes, session key generation, key generation for RSA public key encryption algorithm, and bit stream generation for symmetric stream encryption. There are two fundamentally different strategies for generating random bits. One strategy is to generate each bit based on an unpredictable physical process. This type of random bit generator (RBG) is usually called a non-deterministic random bit generator (NRBG). Another strategy is to use algorithms to calculate bits deterministically. This is called a deterministic random bit generator (DRBG) \cite{barker2012nist}. The DRBG algorithm generates a bit sequence according to an initial value determined by a seed, which is determined by a seed determined according to the output of the randomness source. 
    The seed used to instantiate DRBG must contain enough entropy to ensure randomness. If the seed is kept secret and the algorithm is properly designed, the bits output by DRBG will be unpredictable. We will discuss RBG related instruction set architecture and algorithm research in Section \ref{crypto_algorithm_ISE} and Section \ref{crypto_primitive}, respectively.
    \item \textbf{Trusted Boot Services} reduce the risk of firmware rootkits. It starts with a first-stage Boot ROM that is synthesized into gates. A device with secure boot \cite{liu2017study, lebedev2018secure} enabled will first verify whether the firmware is digitally signed when it starts, and the firmware will check the digital signature of the bootloader to verify that it has not been modified. The bootloader of a trusted boot-enabled device verifies its digital signature before loading the kernel. The kernel sequentially verifies all other components in the boot process, including boot drivers and boot files. If the file has been modified, the bootloader will detect the problem and refuse to load the damaged component. RoT enables trusted servers on the network to verify the integrity of the system boot process. 
    The trusted boot services can be implemented in the RoT, but the service routines are mainly related to system runtime. In this article, we will not discuss trusted boot services in detail. We will discuss them in our next survey of the RISC-V system and application security, which will be a companion to this article.
\end{itemize}
%
%
%
\begin{figure}[t]
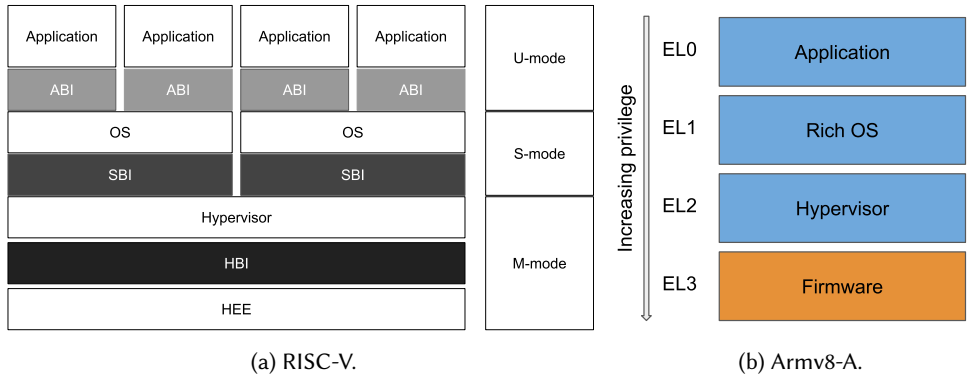
%
    \centering
    \begin{subfigure}{0.5\textwidth}
    \includesvg[height=4.4cm]{figures/privilege_modes.svg}
    \caption{RISC-V.}%
    \end{subfigure} 
    \begin{subfigure}{0.32\textwidth}
    \includesvg[height=4.4cm]{figures/arm_privilege_levels.svg}
    \caption{Armv8-A.}%
    \end{subfigure} 
    \vspace{-2mm}
    \caption{Privileged software stack and corresponding privileged execution modes.}%
\vspace{-4mm}
    \label{riscv_system_stack}%
\end{figure}

\subsection{Building Blocks of RISC-V Architecture Security}\label{building_blocks}
\subsubsection{RISC-V Architecture Stacks and Privilege Modes}
RISC-V can support different software stack implementations. A simple system can be a bare-metal application running on an application execution environment (AEE) in the machine mode (M-mode).   
The application runtime interacts with a particular application binary interface (ABI), which includes the supported user-level ISA and a set of ABI calls to interact with the AEE. The ABI hides details of the AEE from the application, providing an abstract layer for flexibility of implementing the AEE. 

RISC-V can also run an operating system (OS) that can support multiple applications. Each application communicates over an ABI with the OS, which provides the AEE. RISC-V operating systems interface with a supervisor execution environment (SEE) via a supervisor binary interface (SBI). An SBI comprises the user-level and supervisor-level ISA together with a set of SBI function calls. Using a single SBI across all SEE implementations allows a single OS binary image to run on any SEE. The SEE can be a simple bootloader and BIOS-style IO system on a low-end hardware platform, or a virtual machine in a high-end server, or a thin translation layer over a host operating system in an architecture simulation environment \cite{watermanrisc2019}.

As Figure \ref{riscv_system_stack} shows, RISC-V can run a virtual machine monitor configuration where multiple OSs are supported by a hypervisor. This is a typical deployment in complicated infrastructure as a service scenarios. Each OS communicates via an SBI with the hypervisor, which provides the SEE. The hypervisor communicates with the hypervisor execution environment (HEE) using a hypervisor binary interface (HBI) to isolate the hypervisor from the hardware platform.

At any time, a RISC-V hardware thread (hart) is running at some privilege level encoded as a mode. Three RISC-V privilege levels are currently defined: \textbf{User(U) mode} (level 0), \textbf{Supervisor(S) mode} (level 1), and \textbf{Machine(M) mode} (level 3) \cite{watermanrisc2019}. There was a \emph{level 2 H-mode} defined in the RISC-V privileged architecture. H-mode was removed in version 1.10 to enable recursive virtualization support in S-mode. For backward compatibility, the latest RISC-V specification reserves the H-mode. In summary: 

\begin{itemize}
  \item The privilege level is used to provide differentiated protection for different components of the software stack, which lays the foundation for the security of the RISC-V platform.
  \item M-mode is the highest privilege level. It is the only mandatory privilege level of the RISC-V hardware platform. Code running in M mode is usually inherently trustworthy because it has low-level access to the machine implementation. M-mode can be used to manage the secure execution environment on RISC-V.
  \item Many RISC-V implementations support U-mode to protect the rest of the system from application code.
  \item S-mode can be added to provide isolation between a supervisor-level operating system and the SEE.
\end{itemize}

A hart normally runs application code in U-mode until some trap such as a supervisor call or a timer interrupt forces a switch to a trap handler, which usually runs in a more privileged mode. The hart will then execute the trap handler, which will eventually resume execution at or after the original trapped instruction in U-mode. Traps that increase privilege level are termed vertical traps, while traps that remain at the same privilege level are termed horizontal traps. The RISC-V privileged architecture provides flexible routing of traps to different privilege levels.
Each privilege level has a core set of privileged ISA extensions with optional extensions and variants. The M-mode supports an optional standard extension for physical memory protection (PMP) \cite{cheangverifying_pmp}, which is an important security enabler of RISC-V.

The RISC-V privilege level is a similar concept as the ARM exception level. As Figure \ref{riscv_system_stack} shows, the Armv8-A architecture allows implementations to choose whether to implement all exception levels, and select the allowed execution state for each implemented exception level\cite{arm2020levels}. EL0 and EL1 are the exception levels that must be achieved. EL2 and EL3 are optional. The choice not to implement EL3 or EL2 is of great significance. EL3 is the only level that can change the security status. If the implementation chooses not to implement EL3, the PE will not be able to access a single security state. EL2 contains many virtualization functions. No implementation of EL2 can omit these functions. All current Armv8-A implementations support all exception levels \cite{arm2020levels}, because most standard software requires these exception levels. The implementation can also choose the execution state that is valid for each exception level. If AArch32 is allowed at the exception level, it must be allowed at all lower exception levels. For example, if EL3 allows AArch32, it must be allowed at all lower Exception levels. However, existing implementations also have limitations. For example, Cortex-A32 only supports AArch32 for all exception levels. Some modern implementations, such as Cortex-A55, implement all exception levels, but only allow EL0 to use AArch32, exception level EL1, EL2 and EL3 must use AArch64.


\begin{table}[t]
\renewcommand\arraystretch{1.0}
  \centering
  \caption{Comparison of RISC-V PMP and ARM MPU Main Features.}
\vspace{-3mm}
\label{riscvpmp_armmpu}
\addtolength{\tabcolsep}{-1.2pt}
\begin{tabular}{|>{\raggedright}p{5cm}|>{\raggedright}p{5cm}|p{5cm}|}
\hline
{\textbf{}} & {\textbf{RISC-V PMP}} & {\textbf{ARM MPU}}\\
\hline
\hline
\textbf{The smallest region size} & 4 Bytes & 32 Bytes \\
\hline
\textbf{The maximum size of a region}  & 32 GB (if XLEN = 32)& 4 GB  \\
\hline
\textbf{Region granularity}  & Configurable ($2^{G+2}$ Bytes, $G \ge 0$) & 32 Bytes \\
\hline
\textbf{Privileged and unprivileged settings} &  Hybrid (If PMP configuration register L bit is set, the setting also applies to M-mode)  &Independent (Explicitly indicated by the MPU\_RBAR AP field)\\
\hline
\textbf{Supported memory attributes} & R/W/X & R/W/X \\
\hline
\textbf{Maximum number of supported memory regions} & 16 (All for unprivileged, some also applies to privileged if L bit is set) & 16 (8 for privileged, 8 for unprivileged) \\
\hline
\end{tabular}
\vspace{-2mm}
\end{table}

\subsubsection{Physical Memory Protection (PMP)}
For security control, an optional PMP \cite{watermanrisc2019, cheangverifying_pmp} unit provides per-hart machine-mode control registers to allow physical memory access privileges (read, write, execute) to be specified for each physical memory region. The PMP values are checked in parallel with the physical memory attribute checks. In effect, PMP can grant permissions to S and U modes and can revoke permissions from M-mode, which by default has full permissions. PMP violations are always trapped precisely at the processor.
From the viewpoint of functions, RISC-V PMP is the equivalent of ARM MPU \cite{arm-mpu-2016}, which is a programmable unit that allows privileged software to define memory access permissions for separate memory regions. RISC-V PMP and ARM MPU are very similar, but some of their critical configurations such as region size and number of supported regions are different. We compare the key feature set of RISC-V PMP and ARM MPU in Table \ref{riscvpmp_armmpu}. We summarize the main characteristics of RISC-V PMP as follows:
\begin{itemize}
  \item PMP checks are applied to all accesses when the hart is running in \textbf{S} or \textbf{U} modes.
  \item PMP checks are applied to loads and stores when the \emph{MPRV} bit is set in the \emph{mstatus} register and the \emph{MPP} field in the \emph{mstatus} register contains \textbf{S} or \textbf{U}.
  \item PMP checks are applied to page-table accesses for virtual-address translation, for which the effective privilege mode is \textbf{S}.
  \item PMP checks may additionally apply to \textbf{M} mode accesses, in which case the PMP registers themselves are locked, so that even \textbf{M} mode software cannot change them without a system reset.
  \item The standard PMP encoding supports regions as small as four bytes.
  \item RISC-V can maximally support setting sixteen PMP regions.
\end{itemize}

\begin{figure}[t]%
    \centering
    \includesvg[width=12cm]{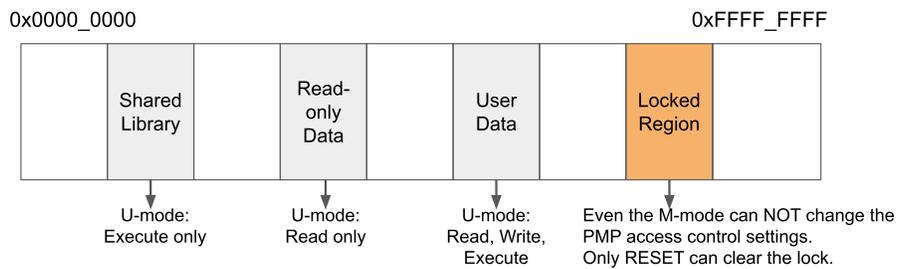}
\vspace{-4mm}
    \caption{Demonstration of RISC-V physical memory protection.}%
\vspace{-2mm}
    \label{riscv_pmp}%
\end{figure}

PMP is mainly used to prevent hart running in lower privilege levels such as U and S modes from accessing privileged memory contents. For example, a regular user hart should be prevented from modifying or even reading the data of shared libraries. Therefore, the memory region in which the shared library data reside can be set as \emph{execute only} using PMP. Even if a hart is running in M-mode, in some cases it is beneficial to prevent it from modifying platform related configurations that may cause runtime errors. By setting the \emph{L} bit of the PMP configuration register, PMP can apply the memory settings to M-mode to enforce protection policies and lock the memory region to prevent the M-mode hart from changing the enforced PMP settings. Figure \ref{riscv_pmp} demonstrates the effect of PMP for runtime memory protection. 

We verified RISC-V PMP settings in M-mode on a \emph{HiFive1 Rev B} board. The reformatted source code is demonstrated as Listing \ref{code:c-code}. The \emph{line\#38} tries to write data into a PMP write-disabled target memory address, which triggers \emph{store/AMO access fault} (exception code 7).   

\begin{code}
\captionof{listing}{Verification of the PMP feature on HiFive1 Rev B board \cite{sifive_boards}. The following codes demonstrate how to configure a PMP Region. The code is adopted from \href{https://github.com/sifive/freedom-e-sdk/tree/master/software/example-pmp}{example-pmp}.}
\vspace{-2mm}
\label{code:c-code}
\begin{minted}[
frame=lines,
%framesep=2mm,
%baselinestretch=1.2,
%bgcolor=LightGray,
fontsize=\footnotesize,
linenos
]
{c}
/* This source code is released under Apache2 and MIT licenses.*/
// Declare Global Variables
int main()
{
	// Declare Local Variables
	/* The "protected_global" is the target PMP memory address
	 * Set the address to be 4-byte aligned and region size to be NAPOT_SIZE bytes */
	size_t protected_addr = ((size_t) &protected_global) >> 2;
	protected_addr &= ~(NAPOT_SIZE >> 3);
	protected_addr |= ((NAPOT_SIZE >> 3) - 1);

	/* Initialize interrupt handling on the current hardware thread */
	cpu = metal_cpu_get(metal_cpu_get_current_hartid());
	cpu_intr = metal_cpu_interrupt_controller(cpu);
	metal_interrupt_init(cpu_intr);

	/* Register the function store_access_fault_handler as the 
	 * handler to process cpu exceptions. */
	rc = metal_cpu_exception_register(cpu, ECODE_STORE_FAULT, store_access_fault_handler);

	/* Reset/initialize the PMP unit */
	pmp = metal_pmp_get_device();
	metal_pmp_init(pmp);

	/* Disable write and execution to protected_global.
	 * The PMP region is locked so it takes effect on M-mode. */
	struct metal_pmp_config config = {
		.L = METAL_PMP_LOCKED,
		/* Set the region's upper bound to be naturally-aligned power of two,
		 * which is determined by the value of A.*/
		.A = METAL_PMP_NAPOT, 
		.X = 0, .W = 0, .R = 1,
	};
	rc = metal_pmp_set_region(pmp, 0, config, protected_addr);

	/* Attempt to write to protected_global. This should trigger a store
	 * access fault exception and enter the registered handler. */
	protected_global[0] = 6;
	
	// Execution shall not arrive at this point if PMP setting is successful
	return 0;
}
\end{minted}
\end{code}


\begin{code}
\captionof{listing}{Demonstration of \emph{AESE} and \emph{AESD} instructions defined in Armv8 Cryptographic Extension \cite{Armv8_2020}. Currently, the standardisation of RISC-V AES and other cryptographic instructions is on-going.}
\vspace{-2mm}
\label{code:aese_aesd}
\begin{minted}[
frame=lines,
%framesep=2mm,
%baselinestretch=1.2,
%bgcolor=LightGray,
fontsize=\footnotesize,
linenos
]
{c}
/* AES single round encryption */
AESE <Vd>.16B, <Vn>.16B
{
    bits(128) operand1 = V[d];
    bits(128) operand2 = V[n];
    bits(128) result;
    result = operand1 EOR operand2;
    result = AESSubBytes(AESShiftRows(result));
    
    V[d] = result; 
}

/* AES single round decryption */
AESD <Vd>.16B, <Vn>.16B
{
    bits(128) operand1 = V[d];
    bits(128) operand2 = V[n];
    bits(128) result;
    result = operand1 EOR operand2;
    result = AESInvSubBytes(AESInvShiftRows(result));
    
    V[d] = result; 
}

\end{minted}
\end{code}

\subsubsection{Cryptographic Instruction Set}\label{crypto_ise}
Securely and efficiently performing cryptographic operations is a basic requirement for a wide range of computing platforms. The dedicated instruction set extension (ISE) is often used to achieve this purpose. 
The implementation of the cryptographic instruction set has two advantages over the software-based implementation. First, the implementation based on CPU hardware can maximize the performance of cryptographic operations. Second, hardware implementation can hide implementation details and reduce the attack surface. The Armv8 Cryptographic Extension provides instructions for the acceleration of AES, SHA, Polynomial Multiply, SM3, and SM4 \cite{Armv8_2020}. Listing \ref{code:aese_aesd} demonstrates the \emph{AESE} (AES Encryption) and \emph{AESD} (AES Decryption) instructions defined in Armv8 Cryptographic Extension. A single execution of these instructions can conduct a round of AES encryption or decryption operation, which actually consists of multiple execution steps.

RISC-V standard cryptographic ISE definition is still ongoing work. There are some active research projects on this topic. Stoffelen et al. \cite{stoffelen2019efficient} present the first optimized assembly implementations of table-based AES, bitsliced AES, ChaCha, and the Keccak-f[1600] permutation for the RV32I instruction set.
Marshall et al. \cite{marshall2020design} further recommend separate ISEs for 32 and 64-bit RISC-V base architectures, with measured performance improvements for an AES-128 block encryption of 4$\times$ and 10$\times$ with a hardware cost of 1.1K and 8.2K gates when compared to a software-only implementation using T-tables. Some RISC-V SoCs such as the PolarFire SoC FPGA have adopted heterogeneous architecture to integrate Arm-based co-processors as the cryptographic engine.
We will further discuss the RISC-V cryptographic ISE research in Section \ref{crypto_algorithm_ISE}.





\subsubsection{Instruction Pipeline}
Instruction pipeline technology divides instructions into a series of sequential steps executed by different processor components, and processes different instructions in parallel in a single processor \cite{olivieri2017investigation}, so that all processor components are fully utilized. Out-of-order execution is a paradigm used in most high-performance processors to take advantage of instruction cycles that would otherwise be wasted. In this paradigm, the processor executes instructions based on the availability of input data and execution units rather than the original order in the program. Thus, the processor can avoid being idle while waiting for the completion of the previous instruction, and at the same time can immediately process the next instruction that can independently run. As a result, out-of-order execution \cite{celio2017boomv2} is an indispensable performance feature of many modern processors. When an out-of-order execution reaches a conditional branch instruction, its direction depends on the instructions that have not yet completed execution. In this case, the processor can checkpoint its current register state, predict the path that the program will follow, and speculatively execute instructions along that path. If the prediction is correct, the checkpoints are not useful, and the instructions are cancelled in the order of program execution. Otherwise, when the processor determines that it follows the wrong path, it will discard all pending instructions along the path by reloading its state from the checkpoint, and resume execution along the correct path to ensure the correctness of the program logic state. The branch predictor tries to guess which way the branch will go before the certainty is known. The purpose of the branch predictor is to improve the flow in the instruction pipeline. In many modern pipelined microprocessor architectures, branch predictors play a vital role in achieving high performance. However, out-of-order execution and speculative branch prediction lead to well-known Meltdown \cite{lipp2018meltdown} and Spectre \cite{kocher2019spectre} vulnerabilities. 

RISC-V ISA avoids over-defining a particular microarchitecture style (e.g., microcoded, in-order, decoupled, out-of-order) or implementation technology (e.g., full-custom, ASIC, FPGA). RISC-V allows efficient implementation in any of these styles \cite{watermanrisc2019}. Rocket Chip \cite{asanovic2016rocket} is an open source Sysem-on-Chip design generator that emits synthesizable RTL. It uses the Chisel hardware construction language to form a complicated generator library for the core, cache, and interconnection to the integrated SoC. Rocket Chip generates general-purpose processor cores, and provides an in-order core generator (Rocket) and an out-of-order core generator (BOOM). Rocket Chip supports the integration of custom accelerators in the form of instruction set extensions, coprocessors or completely independent new cores. Rocket Chip has been taped out and produced a prototype capable of booting Linux. 
Gonzalez et al. \cite{gonzalez2019replicating} replicated Spectre attacks on a BOOM core. To mitigate the attack, they implement a small \emph{L0 speculation buffer} that holds refill data from speculating load misses, and flushes the data when the load is resolved as misspeculated. This prevents misspeculated loads from affecting the state of the cache, while still allowing correctly speculated loads to broadcast their data into the rest of the machine as soon as possible to maintain performance. We will further discuss side-channel prevention related topics in Section \ref{side_channel}. 

In the rest of this article, we will discuss RISC-V hardware and architecture security by topic. Categorizing research papers is challenging, because some research involves multiple topics. For example, when we discuss instruction set extensions, there are related studies on side-channel prevention, but preventing side-channel attack itself is a major research topic, which poses a challenge to the categorization of research papers. In this case, we will discuss a specific study under the topic closest to its work.









\section{Hardware Security}\label{sect_hardware_security}
Embedded devices such as IoT devices face the challenge of physical attacks through side channels or fault injection. Learning from the vulnerabilities of speculative execution, the design of computing architecture should consider security, not just performance. Although the semiconductor industry uses a variety of verification techniques to ensure system-on-chip (SoC) security, attacks are becoming more and more sophisticated. A series of actual attacks that have affected major hardware manufacturers in recent years have proved that ensuring chip security is extremely challenging. It is still a technological trend to reduce software TCB by increasing hardware security extensions. Security solutions such as encryption primitives or TEE based on the underlying TCB will continue to face many challenges in the future. In addition, the security architecture needs to strike an optimal trade-off between the application's high performance and low power consumption \cite{batina2019hardware}. As a new architecture, the RISC-V community is implementing various security solutions. RISC-V's openness and instruction set extension capabilities provide unprecedented opportunities for innovation in the realization of chip security solutions. In this section, we summarize the state-of-the-art RISC-V security research in the hardware and architecture layer. We categorize the existing hardware security research into two topics: hardware and physical security and hardware-assisted security units.

\begin{figure}[t]%
    \centering
    {\includesvg[width=14cm]{figures/Hardware_and_Physical_Security.svg}}
    \caption{RISC-V Hardware and Physical Security. \emph{References: \textbf{Fadiheh2019}\cite{Fadiheh2019}, \textbf{Gleissenthall2019}\cite{gleissenthall2019iodine}, \textbf{Elmohr2020}\cite{elmohr2020fault}, \textbf{Potluri2020}\cite{potluri2020seql}, \textbf{Maas2013}\cite{maas2013phantom}, \textbf{Werner2019}\cite{werner2019protecting}, \textbf{Linscott2018}\cite{linscott2018swan}, \textbf{Takahashi2020}\cite{takahashi2020machine}, \textbf{Bolat2020}\cite{Bolat2020}, \textbf{Dessouky2019}\cite{Dessouky2019Hardfails}.}}%
\vspace{-4mm}
    \label{fig_hardware_physical}%
\end{figure}

\subsection{Hardware and Physical Security} \label{subsect_hardware_physical_security}
Hardware vulnerabilities \cite{fadiheh2019processor} may be caused by unintentional design errors and maliciously implanted hardware Trojans during design or manufacturing. Incorrect design specifications, defective design implementation, or incorrect translation of design in RTL synthesis may all lead to design errors. Recent studies have confirmed that hardware vulnerabilities can be exploited by various attacks, including physical access attacks \cite{maas2013phantom,werner2019protecting}, covert channels \cite{lipp2018meltdown, kocher2019spectre}, fault injection \cite{elmohr2020fault}, and logical lock attacks \cite{Dessouky2019Hardfails}. In addition to leaking critical system or user information, a major risk of hardware vulnerabilities is its irreparability. After tapeout, firmware and software updates will not change the hardware runtime behavior of the chip, which may lead to product recalls. A major challenge in preventing hardware vulnerabilities is that current industry detection methods cannot achieve a good hardware vulnerability detection ratio. Therefore, exploring new hardware security and vulnerability detection technologies is a top priority. Figure \ref{fig_hardware_physical} summarizes the latest RISC-V hardware and physical security research and shows a general view of hardware-related vulnerabilities and attack models.

\subsubsection{RTL Bugs}
Dessouky et al. \cite{Dessouky2019Hardfails} organized an international hackathon to show that the current hardware security verification technology is fundamentally limited and cannot detect RTL bugs that are common in real platforms and can lead to hardware vulnerabilities. Specifically, they injected 31 RTL bugs into two open source RISC-V SoC designs to synthesize different types of common hardware vulnerabilities. These bugs include incorrect privilege escalation, address overlapping, improper write permissions to certain system registers, insecure encryption functions, insecure key storage, and hard-coded passwords. More than 50 teams from all over the world have used formal verification, assertion-based simulation, software-based testing and even manual inspection methods to detect these bugs for months. The industry-leading formal verification technology only detected 15 bugs. The detection ratio of bugs that can lead to secret leakage is very low, which shows that further exploration to advance the technological progress of hardware security verification is extremely important. Sadeghi et al. \cite{sadeghi2021organizing} further summarize the lessons learned in these hardware security
competitions, they observed that the main techniques for detecting hardware vulnerabilities include simulation-driven methodologies, information flow analysis, software-driven simulations, and checking RTL codes using Lint tools. They also envision that fuzzing hardware interfaces will be potentially an effective way to detect hardware bugs.

\subsubsection{Hardware Trojans}
The hardware Trojan (HT) is another major security risk that causes hardware vulnerabilities \cite{xiao2016hardware}. Theoretically, malicious engineers and chip manufacturers can modify the hardware design and implementation to include hardware backdoors so that the attacker can fully control the system. Hardware Trojan detection is an important defense mechanism. The development of defense mechanisms against these dangerous Trojans is relatively lagging behind. Hepp et al. \cite{hepp2021tapeout} designed and integrated four hardware Trojans into a post-quantum encryption-enhanced RISC-V microcontroller. The microcontroller was taped out in September 2020. The impact of these HTs is multifaceted, from simple denial of service to side channel vulnerabilities, and the transmission of sensitive information to external observers. For each HT, they use design tools or simulations to estimate the detectability of these Trojans. Their preliminary observation is that some HTs are easily detected by design tools. However, some HTs that modify the software control flow, causing little interference, are not easy to detect. However, the use of these Trojans that modify the software control flow requires covert modifications to the executable code, which increases the difficulty of using these Trojans to implement attacks in reality. This work provides realistic test equipment for hardware Trojan detection tools.

Linscott et al. \cite{linscott2018swan} propose a novel architecture that maps the security-critical part of the processor design to a one-time programmable LUT-free structure. By analyzing the HDL of the target module, a programmable structure can be automatically generated. By letting the trusted party randomly select a mapping configuration for each chip, the proposed scheme can prevent an attacker from knowing the physical location of the target signal. In addition, they provide a decoy option to map security-critical signals to detect hardware Trojans that hit the decoy. Using this defense method, any Trojan that can analyze the entire configurable structure must use complicated logic functions and take up a large silicon area, which greatly increases the possibility of being detected by security tools. They evaluated the solution on the RISC-V BOOM processor and proved that by providing the ability to map each key signal to 6 different locations on the chip, the proposed scheme can reduce the attack success rate by 99\% with an overhead of only increasing the area by 27\%. 

Side-channel detection \cite{kulah2019spydetector, he2017hardware} is an effective method to discover potential hardware Trojans. It can measure any difference in system power consumption, electromagnetic (EM) emanation, and delayed propagation caused by Trojan insertion or modification in the real design to discover potential threats. However, these methods were evaluated on simple design prototypes such as the AES coprocessor. Moreover, the analysis methods used for these methods are limited by some statistical indicators, such as direct comparison of EM traces or T-test coefficients. Takahashi et al. \cite{takahashi2020machine} propose two new detection methods based on machine learning. The first method is to apply a supervised machine learning algorithm on the original EM trajectory to classify and detect hardware Trojans. Its detection rate is close to 90\%, and the false negative rate is less than 5\%. The second method is based on the outlier/novelty algorithm. This method is combined with the signal processing technology based on T-test, and has better performance. The detection rate is close to 100\%, and the false positive rate is less than 1\%. Takahashi et al. have evaluated the methods on the RISC-V general purpose processor. The area size ratios of the three hardware Trojans in the RISC-V processor are 0.53\%, 0.27\% and 0.1\%, respectively. Although the inserted Trojans are small, the new methods can detect them. 

Existing hardware Trojan research mainly focuses on Trojan attacks in logic circuits. There are still few studies on Trojan attacks in embedded memory. Hoque et al. \cite{hoque2018hardware} discuss a new hardware Trojan for embedded SRAM arrays. They demonstrate various types of Trojan circuits in SRAM including resistive short, bridge, and open in circuit nodes. These Trojans can evade industry-standard post-silicon memory testing and enable target data tampering after deployment. They can cause various malicious influences and have multiple activation conditions, have low overhead in power consumption, performance and stability, and incur negligible silicon area overhead. Bolata et al. \cite{Bolat2020} proposed a RISC-V microprocessor protection architecture against hardware Trojans in memory. The architecture is designed to detect the intrusion of hardware Trojans on the system instruction and data memory. The goal is to detect hardware Trojans that can force the microprocessor to run malicious code or read/write data in unauthorized memory locations. The proposed protection architecture relies on two checkers based on Bloom Filter that monitor the instructions fetched from the instruction memory and the access addresses in the instruction and data memory. They apply the protection architecture to a RISC-V FPGA microprocessor to run a set of software benchmarks for case study.
\subsubsection{Logic-locking Attacks}
Logic locking aims to solve the threat of IP piracy in the semiconductor supply chain. This technology adds a key gate with an input driven by a secret key to hide the internal details of the IP. Only when the programmed key is applied, the conversion is reversed to achieve the original function of the IP. Unfortunately, the existing logic lock function is constantly under attack, and it is difficult to achieve the desired goal. Although current attacks are mainly aimed at combinational circuits, these attacks can be extended to actual sequential circuits through scan-chains. Assuming the scan-input and scan-output ports are controllable and observable by an attacker. The attacker can selectively inject inputs to the scan-input port and analyze the responses from the scan-output port, thus functionally pirating the function of the protected IP module. A secure scan-chain is needed to prevent such attacks. Potluri et al. \cite{potluri2020seql} observed that the flip-flop locking on the scan-input port can obfuscate functional output of the scan-output port. Thus, they propose SeqL, which isolates the functional path from the locked scan path and locks flip-flop inputs to achieve functional output corruption. As a result, SeqL can hide the majority of the scan-correct keys which are functionally correct, thus maximizing the probability that the decrypted key observed by the attackers are functionally incorrect. They validated the effectiveness of the proposed solution on a fully-fledged RISC-V CPU and verified that SeqL can resist a broad range of attacks including SAT, Double-DIP, HackTest, SMT, FALL, Shift-and-Leak, and Multi-cycle attacks.

\subsubsection{Electromagnetic Fault Injection}
Electromagnetic fault injection (EMFI) technology is an important security challenge faced by embedded devices. Elmore et al. \cite{elmohr2020fault} proved through experiments that EMFI enables 320MHz RISC-V processors to skip or mistakenly handle instructions, thus confirming the possibility of attackers using EMFI to conduct widespread attacks. In addition, experimental results on Arm and RISC-V embedded processors show that EMFI attacks are more likely to succeed under lower power supply voltages and higher clock frequencies. They also observed that the exception code is useful for understanding the details of the injected fault, which provided further evidence that the instruction had been corrupted in many cases. Currently, there are still not many countermeasures against EMFI attacks.

\subsubsection{Covert channels}
Covert channels are another risk of sensitive information leakage. Meltdown and Spectre covert-channel vulnerabilities \cite{lipp2018meltdown, kocher2019spectre} have been discovered in advanced processors, which have caused the public to be highly alert to hardware security. These covert channels can leak secret data without any explicit information flow between the secret and the attacker. It is generally believed that these covert channels are inherent in the advanced processor architecture based on speculative and out-of-order execution, and low-end processors do not have such security risks. However, Fadiheh et al. \cite{Fadiheh2019} show that covert channel information leakage is widespread, and they may also appear in average complexity processors with sequential pipelines. They propose a formal method called unique program execution checking (UPEC), which can systematically detect and locate covert-channel vulnerabilities. UPEC employs a formal analysis on the microarchitectural level RTL. UPEC defines the set of all state variables (registers, buffers, flip-flops) belonging to the logic part of the computing system's microarchitecture as the \emph{microarchitectural state variables} of an SoC. It defines the subset of microarchitectural state variables that define the state of program execution at the ISA level excluding the program state that is represented in the program's memory as the \emph{architectural state variables}. It defines the content of memory at a protected location as \emph{secret data}. The UPEC property fails if the system under verification exists a state \emph{soc state}, such that the transition to the next state \emph{soc state'} depends on any \emph{secret data}. Here, \emph{soc state} and \emph{soc state'} are vectors of state variables which include only architectural state variables. If any UPEC property failure is detected, then the design may contain covert channels. The effectiveness of UPEC was explored by targeting different design variants of the open-source RISC-V SoC generator Rocketchip \cite{AsanovictEECS-2016-17}.

Timing channels \cite{biswas2017survey} can also leak sensitive information. Due to complicated fast paths and optimization functions, timing channels are difficult to avoid in modern hardware. A promising way to avoid timing channels is to design and verify conditions under which hardware designs can be executed in constant time. Gleissenthall et al. \cite{gleissenthall2019iodine} propose \emph{IODINE}, an accurate clock and constant time method that can eliminate timing channels in hardware. 
IODINE \cite{gleissenthall2019iodine} defines a syntax to translate Verilog code to intermediate code using its language interpreter \emph{VINTER}. Then it executes the intermediate language for timing channel analysis. For a predefined input vector, the analysis is cycle by cycle. In each cycle, it checks the influence set of each variable.
For different input vectors, it expects that the influence set of any variable in these vectors in each loop is the same, otherwise there may be cases where the execution time of different test vectors is not constant.
In view of the hardware circuit described in Verilog, including a set of sources and sinks and a set of specification assumptions, IODINE allows developers to automatically synthesize proofs to ensure that the hardware executes in a constant time. In other words, under a given usage assumption, the time taken from the source to the sink has nothing to do with the operands, processor flags, and interference from concurrent calculations. By using IODINE, encryption hardware designers can ensure that the hardware execution time is not secret value dependent, so their encryption engine will not leak secret keys. Similarly, CPU designers can ensure that programs (such as cryptographic algorithms) are executed in constant time with the correct structure.
Two real bugs were detected by this methodology, one was in the FPU and the other was in the RSA encryption module. A main contribution of this research is that it proposes a formal methodology to verify the existence of timing side channels in hardware design in a deterministic way.

\subsubsection{Physical Access Attacks}
Attackers with physical access to a chip may directly probe the pin signals to observe sensitive information \cite{wang2017probing}. In order to prevent such attacks, the prior art proposes a secure processor to automatically encrypt and check the integrity of all data outside the processor, including data in DRAM and non-volatile memory. Although the security processor encrypts the memory content, DRAM transfers memory addresses in clear text on the memory bus. Attackers can snoop on the memory bus and observe the physical memory addresses accessed, and then collect sensitive data, such as encryption keys or information about user-level programs. To prevent such information leakage, it is necessary to make the memory address traces indistinguishable. Maas et al. \cite{maas2013phantom} propose a new hardware architecture for effective oblivious computation that can ensure data confidentiality and memory trace obliviousness. In other words, since each memory access of the application program will cause the physical memory address of the accessed data to randomly change, the data access will not produce a fixed address pattern on the physical memory. Therefore, the attacker cannot probe any information about the DRAM locations accessed. 
Also in order to prevent physical access attacks, Werner et al. \cite{werner2019protecting} proposed a method to protect RISC-V processors from fault injection attacks. They demonstrated the protection of control flow through instruction stream encryption and decryption, and the protection of conditional branches by adding redundancy to comparison operations and entangling the comparison result with the encrypted instruction stream.

\begin{figure}[t]%
    \centering
    {\includesvg[width=14cm]{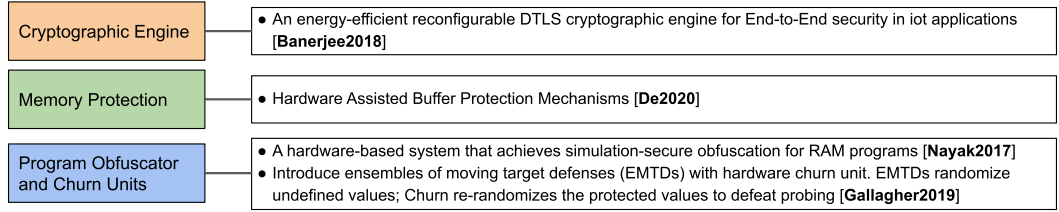}}
    \caption{Hardware-assisted Security Units. \emph{References: \textbf{Nayak2017}\cite{Nayak2017HOP},
    \textbf{Banerjee2018}\cite{banerjee2018energy},
    \textbf{Gallagher2019}\cite{gallagher2019morpheus},
    \textbf{De2020}\cite{de2020hardware}.
    }}%
\vspace{-4mm}
    \label{fig_HardwareExtension}%
\end{figure}

\subsection{Hardware-assisted Security Units}\label{hardware_assist_sec_unit}
Hardware-assisted security units try to ensure a level of security that software cannot provide. The basic assumption of hardware-assisted security mechanisms is that hardware is less likely to have exploitable vulnerabilities than software. The hardware function can reduce the complexity of the software part to improve platform performance. Therefore, the industry is advocating a full range of hardware-assisted security methods including trusted computing, random number generation, crypto acceleration, malware detection \cite{coppolino2019comprehensive}. At the same time, the academic community has also proposed many security solutions based on these industrial hardware trust anchors. The hardware-assisted security unit can be implemented in different forms in many scenarios. For example, in order to achieve the end-to-end security of IoT applications, the hardware unit can be implemented as an encryption engine. In the case of an ensemble of moving target defense, it can be realized as a churn unit. Hardware-assisted units can also be implemented for memory protection and transaction approval. 
RISC-V TEE including \emph{TIMBER-V} \cite{weiser2019timber} and \emph{Keystone} \cite{lee2020keystone} are created based on the memory protection mechanism and trusted boot service. \emph{TIMBER-V} is based on the memory protection unit and tagged memory mechanism, which we will discuss in Section \ref{tagged_mem}. \emph{Keystone} relies on a trusted boot service, which is similar to Arm TEE, and is mainly related to system runtime, which will not be discussed in this article. We plan to discuss TEE as a major topic in our next RISC-V security survey article on systems and applications.
Figure \ref{fig_HardwareExtension} summarizes the research of Hardware-assisted security units that we will discuss in the rest of this subsection.


\subsubsection{Program Obfuscator and Churn Units}
It is challenging to implement virtual black box (VBB) obfuscation of general programs in a pure software manner. Nayak et al. \cite{Nayak2017HOP} proposed \emph{HOP}, which uses secure hardware to realize the simulated security obfuscation of RAM programs. HOP only trusts the hardware processor. The theoretical analysis of HOP considers all the optimizations used in the actual design, including the use of hardware Oblivious RAM (ORAM), hardware scratchpad, instruction scheduling technology, and context switching. They introduced the FPGA prototype hardware implementation of HOP. Through various benchmark evaluations, the cost of HOP is 8 to 76 times that of an insecure system. Compared with all previous efforts to achieve obfuscation (unimplemented), HOP has improved performance by more than three orders of magnitude, making obfuscation technology one major step closer to achieving the goal of being deployable in practice.

Constantly obfuscating the information required by the attacker is an effective counter-attack method. Frequent obfuscation will produce high system overhead. Gallagher et al. \cite{gallagher2019morpheus} proposed \emph{Morpheus}, which is an ensemble of mobile target defense with a hardware churn unit, in which each mobile target defense uses hardware support to provide more randomness at a lower cost. When used in conjunction with obfuscation, Morpheus defense can provide powerful protection against control flow attacks. Security testing and performance research show that Morpheus has achieved high coverage protection against various control flow attacks, including protection against advanced attacks. In addition, a churning period of up to 50 milliseconds is at least 5000 times faster than the time required to penetrate Morpheus.

\subsubsection{Memory Protection}
Code injection and code reuse attacks such as buffer overflow and return-oriented programming (ROP) are still threats to RISC-V programs. De et al. \cite{de2020hardware} proposed two hardware security extensions for RISC-V. First, they use a physical unclonable function (PUF)-based random canary generation technology, which eliminates the need to store sensitive canary words in memory or CPU registers, so it is more secure and efficient. They implemented the proposed Canary engine in RISC-V Rocket Chip. The simulation results show that for a single buffer protection, the average execution overhead is 2.2\%. When the protection is extended to all buffers, increasing the buffer count by 10 times will only increase the overhead by 1.5 times. Second, the author implements Fixer, a dedicated security coprocessor extension for flow integrity. FIXER enforces fine-grained control flow integrity (CFI) for programs running on the backward edge (return) and forward edge (call) without requiring any architectural changes to the processor core. Compared with software-based solutions, FIXER reduces energy consumption by 60\% with minimal execution time (1.5\%) and area (2.9\%) overhead.

\subsubsection{Cryptographic Engines}
Datagram Transport Layer Security (DTLS) is an important protocol for end-to-end IoT communication security. The high computational overhead makes pure software DTLS implementation too costly for resource-constrained embedded devices. Banerjee et al. \cite{banerjee2018energy, banerjee2019energy} demonstrate the first hardware implementation of the DTLS protocol. The key component of the design is the reconfigurable element field elliptic curve encryption (ECC) accelerator, which is 238 times and 9 times more energy efficient than software and the latest hardware implementations. The complete hardware implementation of the DTLS 1.3 protocol is 438 times more energy-efficient than software, and the code size and data memory footprint are as low as 8KB and 3KB, respectively. Benchmarking of applications other than DTLS shows that the combination of a cryptographic accelerator and an on-chip low-power RISC-V processor can save up to two orders of magnitude of energy. Their test chip is made of 65nm CMOS. At 16MHz and 0.8V, each handshake consumes 44.08 µJ and each byte of encrypted data consumes 0.89 nJ.

Co-design of software and hardware can significantly improve the performance of cryptographic algorithms. Wang et al. \cite{wang2019xmss} proposed the software and hardware co-design of the hash-based post-quantum signature scheme XMSS on the RISC-V embedded processor. They provide software optimizations for the SHA-256 parameter set and the XMSS reference implementation of multiple hardware accelerators, allowing area usage and performance to be balanced according to individual needs. Compared with pure software implementation, by integrating the hardware accelerator into the RISC-V processor, the key pair generation performance can be improved by more than 54 times. The signature generation time is less than 10 milliseconds, and the verification time is less than 6 milliseconds, which is 42 times and 17 times faster than software. They tested and measured the number of cycles on Intel Cyclone V SoC FPGA. The integration test of their XMSS accelerator and embedded RISC-V processor shows that the hash-based post-quantum signature can be actually used in a variety of embedded applications.






\begin{figure}[t]%
    \centering
    {\includesvg[width=15cm]{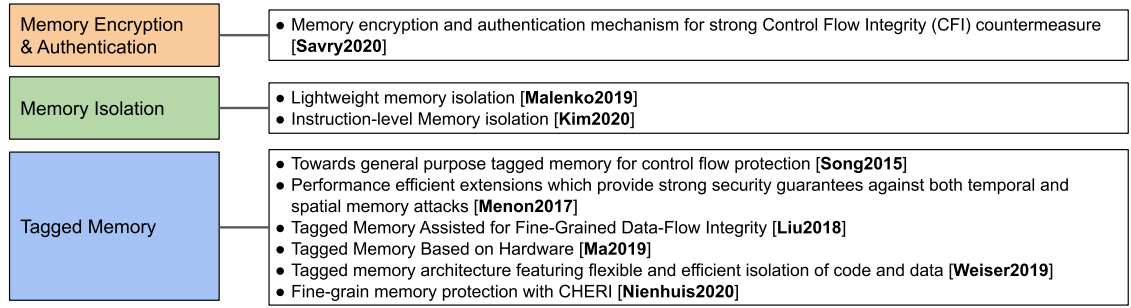}}
    \caption{RISC-V Memory Protection. \emph{References: \textbf{Malenko2019}\cite{malenko2019device}, \textbf{Kim2020}\cite{Kim2020}, \textbf{Savry2020}\cite{Savry2020}, \textbf{Song2015}\cite{Song2015}, \textbf{Menon2017}\cite{menon2017shakti}, \textbf{Liu2018}\cite{Liu2018}, \textbf{Ma2019}\cite{Ma2019Dam}, \textbf{Weiser2019}\cite{weiser2019timber}, \textbf{Nienhuis2020}\cite{nienhuis2020rigorous}}}%
\vspace{-4mm}
    \label{fig_memory_protect}%
\end{figure}

\section{Memory Protection}\label{mem_protect}
Retrospecting the history of computer engineering, many security vulnerabilities originated from two aspects. First, mainstream processor architectures and C/C++ language abstractions have only provided coarse-grained virtual memory-based protection since the 1970s. Second, the mainstream engineering methodology follows the process of design, development, testing, and debugging. This methodology can satisfy many areas of the computer industry, but they fundamentally leave a large number of exploitable vulnerabilities, leading to many serious system security problems \cite{nienhuis2020rigorous}. Effective memory protection can alleviate many system security vulnerabilities. As summarized in Figure \ref{fig_memory_protect}, At least three types of memory protection solutions have been proposed: tagged memory\cite{Song2015,menon2017shakti,Liu2018,Ma2019Dam,weiser2019timber,nienhuis2020rigorous}, memory isolation \cite{malenko2019device,Kim2020}, and memory encryption and authentication\cite{Savry2020}. In this chapter, we discuss related research work on these topics. 

\subsection{Tagged Memory} \label{tagged_mem}
The tagged memory \cite{zeldovich2008hardware, joannou2017efficient} enables a memory pointer to have a capability tag. Upon a memory access, the tag of the memory address will be checked to see if there is any capability violation. Tagged memory can prevent typical information leakage that may easily happen in classic programming languages. For example, the ISO C program can leak the secret key data because of an unintentional pointer access. Even worse, the compiler will not report any warning to this. With tagged memory such as its implementation in the Cheri C language extension \cite{watson2020cheri}, this kind of bug cannot happen because the capability checking mechanism will block unsafe memory access. Memory protection features have also been supported in some emerging programming languages such as the RUST, which is getting popular because of its security features and efficiency. Tagged memory is an active research domain of RISC-V hardware and architecture. 

Song et al. \cite{Song2015} explored the performance of tagged memory by extending the Rocket RISC-V implementation \cite{AsanovictEECS-2016-17} with preliminary tagged memory support. The implementation adds a predefined number of tag bits to each 64-bit word in memory. These tag bits are copied along with the data word through the cache hierarchy, meaning that each word in the L1 data and L2 cache lines are augmented with additional tag bits. The coherence of tags is maintained by the existing cache coherence mechanisms. Two new instructions \emph{LTAG} and \emph{STAG} are added for loading and storing tags. The tags are stored in a reserved memory area. Each access to a memory word also needs access to the tag, resulting in a memory traffic ratio of 2. Employing a tag cache can reduce the memory traffic ratio, and the traffic reduction is tag cache size dependent. Increasing the cache size from 16KB to 128KB can reduce the average memory traffic ratio from 1.59 to 1.06.
Data-oriented attacks \cite{cheng2019exploitation} manipulate non-control data to alter a program's benign behavior without violating its control flow integrity. It has been shown that such attacks can cause significant damage even in the presence of control-flow defense mechanisms.
Based on the tagged memory feature, Liu \cite{Liu2018} and Ma \cite{Ma2019Dam} et al. present tagged memory supported data-flow integrity mechanisms to enable fine-grained data-flow integrity checking to mitigate data-oriented attacks.

Many memory vulnerabilities are related to pointers \cite{gens2018k}. Spatial memory attacks occur when a particular pointer accesses memory regions beyond its permissible range. Temporal memory attacks, on the other hand, occur when accessing a memory region that has been freed after allocation. Tagged memory is an effective scheme for memory pointer protection. Menon et al. \cite{menon2017shakti} propose on-chip hardware design extension called the Base-and-Bound Cache (BnBCache), which optimizes fat-pointer performance via reducing the total number of memory accesses. It assumes that each 64-bit word in the memory is associated with a single Tag-Bit indicating whether the word is a pointer or regular data. The Tag-bits are set by the compiler and stored alongside the memory word even when it is loaded into a register, which also supports Tag-Bit. 
\emph{BnBCache} consists of a \emph{BnBIndex} table and a \emph{BnBLookUp} table. The entries in the BnBIndex table have a 1 to 1 mapping with 32 general purpose registers. Each BnBIndex entry has an index pointing to a BnBLookUp entry, which consists of 4 fields: the base value (64-bits), the bound value (64-bits), the ptr\_id (64- bits) and a valid bit. ISA extensions with eight new instructions are proposed to support the tagged memory mechanism. With the tag bit, valid bit, and boundary information, both spatial and temporal memory attacks can be blocked. The proposed solution, which is implemented on top of a RISC-V ISA based 64-bit baseline processor, incurs an area overhead of 1914 LUTs and 2197 flip flops on an FPGA without bringing critical path delay \cite{menon2017shakti}.  

\begin{figure*}[t]
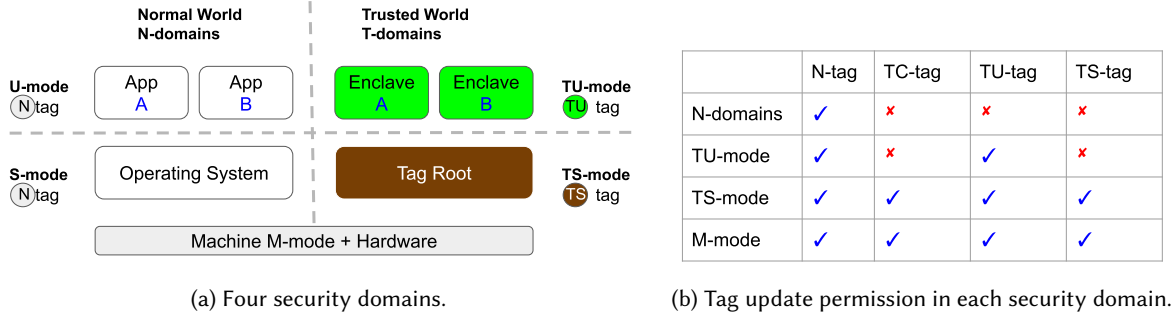
%
    \centering
    \begin{subfigure}[b]{0.55\textwidth}
         \centering
         \includesvg[width=8.5cm]{figures/timber-v-fig1.svg}
         \caption{Four security domains.}
         \label{fig:domain}
     \end{subfigure}
     \hfill
     \begin{subfigure}[b]{0.43\textwidth}
         \centering
         \includesvg[width=6.5cm]{figures/timber-v-fig2.svg}
         \caption{Tag update permission in each security domain.}
         \label{fig:tag_update}
     \end{subfigure}
\vspace{-4mm}
    \caption{TIMBER-V Fine-grained Enclaves based on Tag-Isolated Memory. This figure is adopted from paper \cite{weiser2019timber}.}%
\vspace{-4mm}
    \label{fig_timber-v}%
\end{figure*}

TIMBER-V \cite{weiser2019timber} is a new tagged memory architecture. Combined with the memory protection mechanism, it can flexibly and efficiently isolate code and data to implement a Trusted Execution Environment (TEE) on small embedded systems. As Figure \ref{fig:domain} demonstrates, execution in user mode (U-mode) and supervisor mode (S-mode) can both be separated in the normal world and the trusted world. The N-domains in the normal world support the traditional split between U-mode and S-mode, and allow existing code to run without modifications. Memory words in N-domains, no matter under U-mode or S-mode are encoded with the \emph{N} tag. Memory words in T-domains under U-mode and S-mode are encoded with the \emph{TU} and \emph{TS} tags, respectively. Trusted user mode (TU-mode) can be leveraged for isolated execution environments, called enclaves. Trusted supervisor mode (TS-mode) allows to run TagRoot trust manager, augmenting the untrusted operating system with trusted services.
Switching from N-domains to T-domains is implemented via trusted callable entry point functions, which is encoded with the \emph{TC} tag. There are totally 4 different tags. Thus, TIMBER-V uses a two-bit tag per 32-bit memory word. Tags can only be updated within the same or a lower security domain but cannot be used to elevate privileges, as shown in Figure \ref{fig:tag_update}. TS-mode (and M-mode) have full access to all tags. TU-mode can only change tags between N-tag and TU-tag to support dynamic interleaving of user memory. TU-mode is prevented from manipulating TC-tags, which are reserved for secure entry points.
TIMBER-V uses the MPU to enhance the label isolation function to isolate each process while maintaining low memory overhead. 
TIMBER-V greatly reduces memory fragmentation and improves the dynamic reuse of untrusted memory across security boundaries. In addition to interleaving stacks, TIMBER-V can also implement novel execution stack sharing across different security domains. TIMBER-V is compatible with existing code, supports real-time constraints. A proof-of-concept implementation of TIMBER-V has been evaluated on the RISC-V simulator.
Similar to the TIMBER-V, CHERI architecture also provides hardware capabilities that supports fine-grained memory protection and scalable secure compartmentalisation. Porting the tagged memory feature of CHERI to the RISC-V is on-going \cite{nienhuis2020rigorous}.

\subsection{Memory Isolation, Encryption and Authentication}
Existing memory isolation mechanisms suffer from scalability and performance issues, Kim et al. \cite{Kim2020} propose instruction-level memory isolation (RIMI) to boost performance. RIMI also introduces the concept of domain to separate code and data in different memory map areas. The domain memory protection (DMP) mechanism only allows domain specific instructions to access instruction and data in a corresponding domain to achieve instruction-level memory isolation of different domain accesses. Each domain also consists of physical memory protection (PMP) regions and dedicated instructions to access its PMP regions. PMP and DMP configurations are jointly checked to determine the access permission of instruction and data.
Specifically, RIMI implements memory \emph{load} and \emph{store} instructions with \emph{domain\_id} tags, and special control transfer instructions with \emph{x} tags for domain switching. For example, \emph{lw1} indicates the memory load instruction to access domain1, \emph{sw0} indicates the memory storage instruction to access domain0, \emph{jalr} can only jump between the same domain, and \emph{jalrx} can jump between different domains.
Evaluation based on the Spike simulator shows that using RIMI can effectively achieve shadow stack and in-process isolation.
Also at the instruction level, Savry et al. \cite{Savry2020} proposed a framework to ensure control flow integrity by implementing a lightweight masking scheme applied to instructions, which is based on an authenticated memory encryption mechanism. At the system level, for memory isolation, Malenko et al. \cite{malenko2019device} implemented a device driver isolation module in RTOS to prevent defective device drivers from destroying the state of the operating system and applications.











\begin{figure}[t]%
    \centering
    {\includesvg[width=14cm]{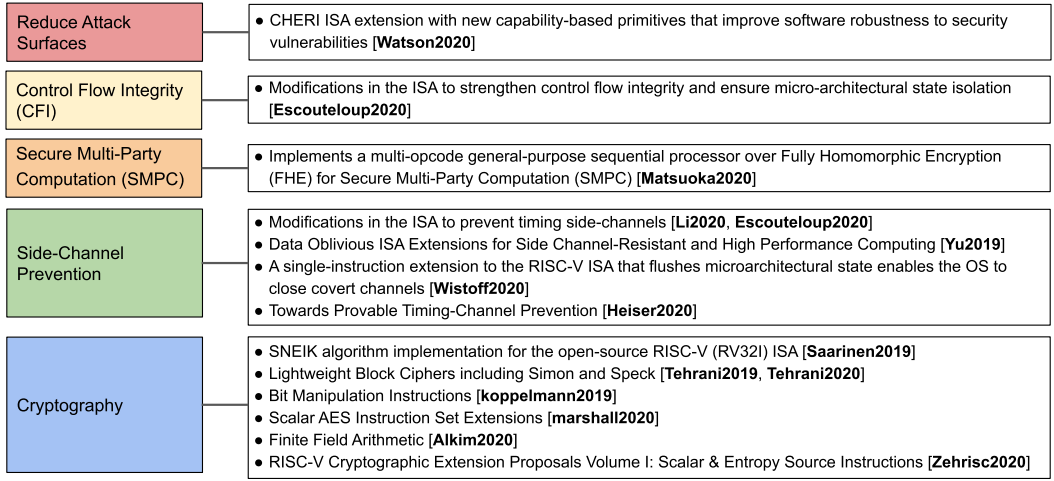}}
    \caption{RISC-V ISA Extensions for Various Security Objectives. \emph{References: \textbf{Li2020}\cite{Li2020SIMF}, \textbf{Escouteloup2020}\cite{escouteloup2020recommendations}, \textbf{Yu2019}\cite{yu2019data}, \textbf{Wistoff2020}\cite{wistoff2020prevention}, \textbf{Heiser2020}\cite{heiser2020towards}, \textbf{Watson2020}\cite{watson2020capability}, \textbf{Saarinen2019}\cite{saarinen2019sneik}, \textbf{Tehrani2019}\cite{tehrani2019classification}, \textbf{Tehrani2020}\cite{Tehrani2020DSD} , 
    \textbf{Koppelmann2019}\cite{koppelmann2019risc}, \textbf{Alkim2020}\cite{alkim2020isa}, \textbf{Marshall2020}\cite{marshall2020design}, \textbf{Zehrisc2020}\cite{zehrisc2020}, \textbf{Matsuoka2020}\cite{Matsuoka2020Virtual}.}}%
\vspace{-2mm}
    \label{fig_ISA_Extension}%
\end{figure}

\section{ISA Security Extensions}\label{ISA_Security_Extensions}
The instruction set architecture (ISA) unifies the behavior of machine code running on the different CPU implementations of the ISA. The life cycle of ISA spans decades, but applications usually evolve quickly. In order to adapt to new application requirements, ISA may need to add new features to optimize performance, energy efficiency, or security. The instruction set extension is usually used to achieve the above goals while maintaining backward compatibility. Well-known instruction set extensions include x86 FPU, SSE, AVX, AES, SGX, and Arm SVE, Thumb, Neon, VFPv4 and TrustZone security extensions. As summarized in the figure \ref{fig_ISA_Extension}, related extensions of RISC-V ISA have been proposed to enable hardware cryptographic functions \cite{saarinen2019sneik,Tehrani2020DSD,koppelmann2019risc,marshall2020design,alkim2020isa,zehrisc2020}, reduce system attack surfaces \cite{watson2020capability}, resist certain side-channel attacks \cite{Li2020SIMF,escouteloup2020recommendations,yu2019data,wistoff2020prevention,heiser2020towards}, strengthen control flow integrity \cite{escouteloup2020recommendations}, and achieve secure multi-party computation \cite{Matsuoka2020Virtual}. We discuss RISC-V ISA security extensions in this Section.

\subsection{Cryptographic Algorithms}\label{crypto_algorithm_ISE}
Cryptographic algorithms are ubiquitous in platform security mechanisms. These algorithms are compute-intensive. There exist cryptographic standards and guidelines for a wide range and cryptographic functions including block cipher techniques, digital signatures, hash functions, and key management etc. Performance requirements and standard implementations of cryptographic functions make it a proper solution to integrate these functions as ISA extensions and implement them in hardware. We have demonstrated Armv8 Cryptographic Extension for \emph{AESE} and \emph{AESD} instructions in Section \ref{crypto_ise}. The RISC-V community is also extending the instruction set for similar purposes. 

Secure and efficient implementation of AES is a basic requirement on most computing platforms. Therefore, dedicated instruction set extensions (ISEs) are often implemented to support efficient AES execution. RISC-V is a new ISA and lacks this standardized ISE. Marshall et al. \cite{marshall2020design} investigated the latest industrial and academic ISEs for AES, and evaluated five different ISEs. Compared with the software T-table based implementation, ISEs for 32-bit and 64-bit architecture can achieve 4$\times$ and 10$\times$ AES-128 block encryption performance improvement with hardware costs of 1.1K and 8.2K gates, respectively. They also explored how to use RISC-V standard bit manipulation extension \cite{wolfrisc2021} to effectively implement AES-GCM. Their work is part of the ongoing RISC-V cryptography extension standardization process. 
RISC-V Cryptographic Extensions Task Group are working on cryptographic extension proposals for scalar and entropy source instructions, which covers bit manipulation, scalar AES, SHA, SM3 and SM4 acceleration, and TRNG entropy source interface \cite{zehrisc2020}.  

Similar to the AES GCM algorithm, SNEIK \cite{hashing2019sneiken} is a lightweight, permutation-based encryption primitive code library that can perform cryptographic hashing, authenticated encryption with associated data, and other tasks. The design is designed to meet all symmetric cryptographic needs, including tasks such as pseudo-random number generation and key derivation. 
Saarinen et al. \cite{saarinen2019sneik} evaluated SNEIK on RISC-V (RV32I) and showed that SNEIKEN128 can conduct authenticated encryption at 54.8 instructions/byte, which roughly matches the performance of AES-128 on the comparable Arm platform. SNEIKHA256 achieves 98.6 instructions/byte on RV32I. The RV32I base instruction set lacks rotation instructions, which results in lower throughput than Armv7. They observed that the structure of SNEIK permutation was very suitable for ISA extension optimization. RV32I extension only has an impact of 258 LUT / 65 slices on FPGA resource utilization, but it increases the SNEIK permutation speed  by 7$\times$. Tests conducted on Artix-7 FPGA hardware showed that the RISC-V ``Crimson Puppy'' SoC with ISA extension can perform SNEIKEN128 operation at 12.4 cycles/byte, and SNEIKHA256 operation at 17.3 cycles/byte, demonstrating that a simple RISC-V instruction set extension can achieve 5$\times$ acceleration. 

Tehrani et al. \cite{tehrani2019classification, Tehrani2020DSD} provides a detailed architecture and implementation of specific processor instructions for lightweight encryption algorithms. These instructions target the 32-bit RISC-V ISA and allow acceleration of several commonly used lightweight block ciphers. They used the plug-in-based architecture of the VexRiscv processor to implement instruction extension on the Artix-7 FPGA board. They demonstrated the hardware resource usage of these extended instructions. For a representative lightweight block cipher, they compared the performance of the system with ISA extensions and the base system. The results show that the instruction extension can accelerate the lightweight encryption algorithm by 33 to 138$\times$ at a reasonable hardware cost. We will further discuss lightweight cryptographic algorithms in Section \ref{crypto_primitive}.

Bit manipulation is the act of processing bits shorter than words. Cryptographic algorithms require a large number of bit operations, so the support of bit operations has a significant impact on the performance of cryptographic algorithms. Koppelmann et al. \cite{koppelmann2019risc} proposed RISC-V extensions for bit manipulation instructions (BMIs). Specifically, they extended the RISC-V ISA with ten bit manipulation instructions: parity, byte swap, right/left rotation, popcount, bit reverse, count leading/trailing zeros, and parallel gather/scatter. These BMIs achieve the same functions as the current x86 BMIs, while the required code bytes are reduced by 13.5\%. In order to prove its efficiency, they evaluated the extensions using 13 benchmarks, and the new instructions showed good acceleration.
Wolf et al. \cite{wolfrisc2021} in RISC-V BitManip Task Group is actively working on the bit manipulation instruction extensions. According to the RISC-V Bitmanip Extension Document Version 0.94-draft \cite{wolfrisc2021}, they have proposed more than 100 bit manipulation instructions for both 32 and 64-bit ISAs, covering bit manipulating, bit permutation, bit field place, bit compress/decompress etc. 

The development of quantum computing and the devastating impact of Shor's algorithm on our current IT security has spawned active research on encryption systems that prevent quantum computing attacks. This field of research is called post-quantum cryptography (PQC). An important aspect of PQC research is the effective and secure implementation of PQC algorithms. Current cryptographic algorithms require effective arithmetic operations on hundreds to thousands of bits of data, while many PQC schemes perform finite field operations on data less than 20 bits. Alkim et al. \cite{alkim2020isa} take the lattice-based key encapsulation mechanisms Kyber and NewHope as examples to study the impact of providing ISA extensions with finite field operation support on the performance of the PQC schemes. They create a prototype implementation of the presented instructions on the VexRiscv core, and evaluate the design on two different FPGA platforms. The result shows a speedup of polynomial arithmetic of up to 85\% over the basic software implementation. The custom instructions can replace a general purpose multiplier to achieve very compact implementations.

Fault attacks and power analysis threaten the implementation of cryptographic schemes. Shielded power analysis and redundancy-based methods are the most commonly used countermeasures for this type of attack. With these attacks in mind, NIST recently requested the submission of documents to illustrate the possibility of adding countermeasures against these attacks at low cost. Steinger et al. \cite{steinegger2020fast} propose an instruction extension of Ascon-p, which uses tight integration with the processor register file to significantly accelerate various symmetric encryption calculations at a relatively low cost. As a proof of concept, they integrated the instruction extension into the 32-bit RI5CY core. They evaluated various hardware indicators and showed that the accelerator can be implemented with about 4.7 kGE, in other words about half the area of the dedicated coprocessor design. Considering this built-in acceleration of Ascon-p, they created an assembly version of the Ascon/Isap model, which utilizes instruction extensions and provides benchmarks for authentication encryption, hashing, and pseudo-random number generation. Based on the benchmarks, compared with the pure software implementation, instruction extensions of Ascon-p achieved a speedup of 50 to 80.


\begin{figure}[t]%
    \centering
    {\includesvg[width=8cm]{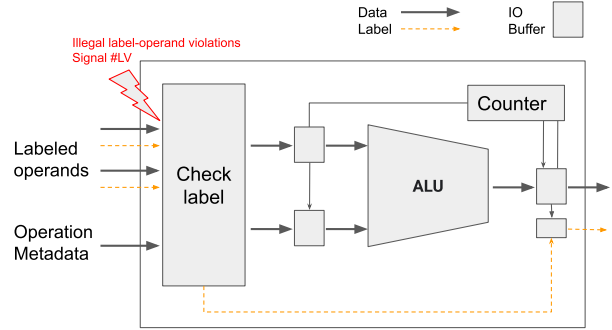}}
    \caption{Label station for an execution unit with one internal arithmetic unit in OISA architecture \cite{yu2019data}.}%
\vspace{-4mm}
    \label{fig_yu2019}%
\end{figure}

\subsection{Side-channel Attack Prevention}\label{isa_side_channel_prevent}
As we discussed in Section \ref{subsect_hardware_physical_security}, preventing microarchitecture side-channel attacks is one of the most pressing challenges in hardware security today. Meltdown\cite{lipp2018meltdown} and Spectre\cite{kocher2019spectre} belong to this type of attack. In order to prevent side-channel attacks, hardware-based technology focuses on redesigning the cache and completely modifying the processor architecture to improve cross-processor information flow tracking. The software-based method recommends clearing all core states, including private caches, translation backup buffers, branch prediction units, etc. The software-based method relies heavily on the refresh mechanism. Although refreshing or clearing the persistent state of multiple vulnerable hardware components at the core level (within level 1) is essential to create good and complete time isolation in the entire system, research shows that the existing ISA refresh instructions are functionally incomplete and inefficient in performance and power efficiency. 

There has been related work trying to prevent side-channel information leakage by obliviously writing program data. In this model, program writing needs to avoid sensitive data access leaving traces on shared resources. Despite recent efforts, the security and performance of running data-oblivious programs on modern computers are questionable. First, writing a data-oblivious program assumes that certain instructions in the ISA will not leak information, but the ISA and hardware do not provide such guarantees. Second, writing programs to avoid data-dependent behavior will inevitably incur serious performance overhead.
Utilizing ISA extension to prevent side-channel attacks has been proposed \cite{yu2019data, escouteloup2020recommendations, Li2020SIMF, wistoff2020prevention, heiser2020towards}. In this section, we discuss in detail the related work of preventing side-channel attacks through RISC-V instruction extension.

Yu et al. \cite{yu2019data} proposed a data oblivious ISA extension (OISA) for RISC-V. In terms of security, the proposed ISA design can block side channels. In terms of performance, the OISA supports effective memory oblivious calculations, and has security features that can remain enabled in common situations such as out-of-order speculative execution and other modern hardware optimizations. Based on the RISC-V out-of-order, speculative BOOM processor, they implemented a complete hardware prototype. Through formal analysis of an abstract BOOM-style machine, they proved that OISA can achieve its security design goals. They evaluated the area overhead of the hardware mechanism, and provided performance experiments to show how OISA can improve various existing data oblivious codes, including constant-time cryptography and memory oblivious data structures, while also improving their security and portability. The label station is the OISA core component. As Figure \ref{fig_yu2019} demonstrates, the label station checks and tracks Public/Confidential labels as data flows through the pipeline and signals \emph{\#LV} when violations occur. The result label, which travels with the result and accompanies the entire life cycle of the result, is computed based on operand labels.

We have discussed timing channels in Section \ref{subsect_hardware_physical_security}. ISA extension is also a potential way to prevent timing channels. Li et al. \cite{Li2020SIMF} created a dedicated flushing instruction to improve the efficiency of temporary isolation at the core level to mitigate the possibility of potential timing channels. They first propose a single instruction multiple refresh (SIMF) scheme, which integrates the refresh operation in one instruction to clear the core-level state. The main advantages of SIMF are: 1) It greatly reduces the dynamic instruction count dedicated to refresh (resulting in cycle counting and instruction fetching capabilities); 2) It requires minimal expansion of existing hardware (adding an instruction to the ISA); 3) When SIMF is not in use, in the case of explicit barrier instructions, the sequence of refresh operations is implicitly executed in an instruction; 4) It brings benefits for programming, including atomicity and simplicity. They prototyped SIMF in an open source scalar ordered RISC-V processor. They extended the RISC-V ISA with another instruction called \emph{FLUSHX}, which refreshes the core-level state, including L1/L2 TLB, L1 cache, and branch prediction units (BTB, RAS, BHT). Wistoff et al. \cite{wistoff2020prevention} proposed a similar solution targeting the similar timing channel issue. Specifically, they propose a new RISC-V fence instruction with arguments to enable the operating system to control state flushing. The evaluation shows that the proposed scheme completely eliminates the timing channels including L1 data and instruction cache channels, TLB, branch target buffer (BTB), and branch history table (BHT) channels.

\subsection{SMPC and CFI}
The goal of secure multi-party computation (SMPC) is to create methods for all parties to jointly compute functions on their inputs, while keeping those inputs private \cite{goldreich1998secure}. Traditionally, cryptography is about hiding content, and this new type of calculation and protocol is about hiding part of the information about the data while using data from multiple sources for calculations and correctly generating output. By customizing RISC-V ISA, Matsuoka et al. \cite{Matsuoka2020Virtual} have proposed the Virtual Security Platform (VSP), which is a comprehensive platform that can provide a complete set of tools for a complete two-party secure computation offloading (SCO) solution. The VSP includes open source design and implementation of homomorphic encryption library, processor architecture, custom ISA and compiler environments. Based on the famous Torus Fully Homomorphic Encryption (TFHE) scheme, VSP allows any user with any C program to execute its code in SCO mode.

The goal of Control Flow Integrity (CFI) \cite{abadi2009control} is to prevent malware from attacking the control flow of the redirector. We have discussed hardware-assisted security units to achieve CFI in Section \ref{hardware_assist_sec_unit}. CFI can also be enforced through ISA extensions. Escouteloup et al. \cite{escouteloup2020recommendations} discuss changes to the ISA to strengthen CFI and ensure the isolation of microstructure states. They put forward some security recommendations in the ISA design, such as marking certain registers such as the frame pointer as confidential to disable branching on confidential registers, authorizing instructions only when the timing of the instruction is not data-dependent, and prohibiting forward indirection jump and prohibit all micro-architectural management instructions, especially cache management, and must provide micro-architectural security guarantees through hardware security context (HSC) instructions.

\subsection{Reduce Attack Surface}
The attack surface of the system is defined as the attackability of the system in the three abstract dimensions of method, data and channel \cite{manadhata2010attack}. Intuitively, the larger the attack surface, the more likely the system is to be attacked, and therefore the less secure it is.
Capability Hardware Enhanced RISC Instructions (CHERI) \cite{watson2020capability} extends the instruction set architecture (ISA) with new function-based primitives, thereby improving the robustness of the software in terms of security. The CHERI model follows the principle of least privilege, and achieves higher security by minimizing the privileges that can be accessed by running software. Another guiding principle that CHERI adheres to is the principle of intentional use, that is, when a software can use many privileges, the use of privileges should be clearly specified, rather than an implicit choice. CHERI reduces the attack surface of the system, and even if an attacker successfully exploits the vulnerability, they will gain the least permissions. CHERI has previously been applied to MIPS and Arm ISAs. Recently, CHERI and its complete software stack has been ported to the RISC-V 32-bit and 64-bit variants \cite{watson2020capability}. CHERI-RISC-V specification shares a lot of architectural features with the CHERI-MIPS such as the tagged memory. But they have different interpretations of addresses in memory capabilities. Also, CHERI-related page permissions are added to RISC-V architectural page-table formats instead of the MIPS translation lookaside buffer (TLB) entries.












\begin{figure}[t]%
    \centering
    {\includesvg[width=14cm]{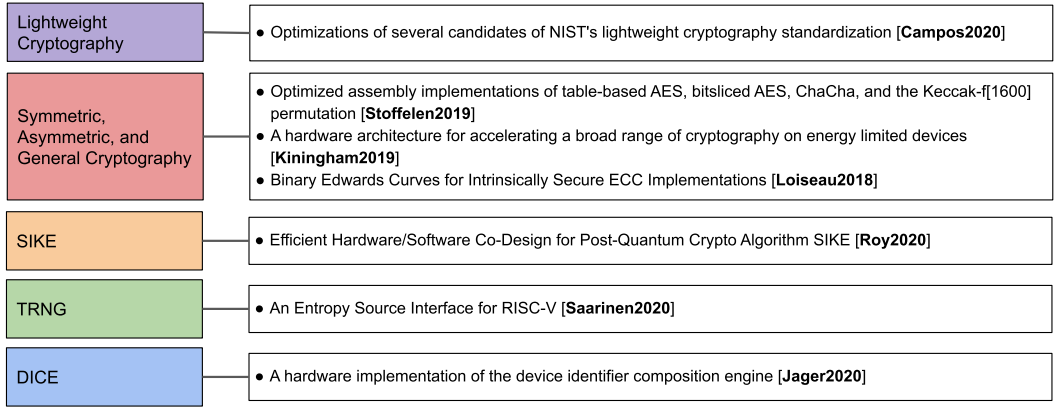}}
    \caption{Cryptographic Primitives. \emph{References: \textbf{Stoffelen2019}\cite{stoffelen2019efficient}, \textbf{Kiningham2019}\cite{Kiningham2019Falcon}, \textbf{Campos2020}\cite{Campos2020}, \textbf{Roy2020}\cite{roy2020efficient}, \textbf{Saarinen2020}\cite{saarinen2020building}, \textbf{Jager2020}\cite{Jager2020DICE}, \textbf{Loiseau2018}\cite{loiseau2018binary}}.}%
\vspace{-2mm}
    \label{fig_crypto_primitives}%
\end{figure}

\begin{table*}[t]
\renewcommand\arraystretch{1.0}
  \centering
  \caption{Performance of Polarfire RISC-V FPGA vs. Xeon x86 Data Center-level CPU Cores. The evaluation is based on openssl cryptography implementation with a single thread. The metrics are KB per second processed for HMAC, SHA, and AES, operations per second for ECDH, and sign or verify operations per second for RSA and ECDSA.}
\vspace{-3mm}
\label{openssl_perf}
\addtolength{\tabcolsep}{-1.2pt}
\begin{tabular}{|>{\raggedright}p{2.8cm}|c|c|c|c|c|c|c|c|}
\hline
  \multirow{2}{*}{\textbf{
\hspace{-2.0mm}
\hspace{-2.0mm}}} & \multirow{2}{*}{\textbf{HMAC (MD5)}} & \multirow{2}{*}{\textbf{\hspace{-1.0mm}SHA256\hspace{-1.0mm}}} &\multirow{2}{*}{\textbf{\hspace{-1.0mm}AES-256 CBC\hspace{-1.0mm}}} &\multirow{2}{*}{\textbf{ECDH}}& \multicolumn{2}{c|}{\textbf{RSA}} &\multicolumn{2}{c|}{\textbf{ECDSA}} \\
  \cline{6-9}
&&&&&\textbf{Sign}&	\textbf{Verify}&\textbf{Sign}&	\textbf{Verify}\\
\hline
\hline
\textbf{Polarfire RISC-V FPGA (660MHz)}&	32421&	7392&	6081&	155&	3693&	54.4&	84&	57\\
\hline 
\textbf{Xeon x86 Server (2.5GHz)}&	572497&	386888&	173803&	11576&	176056&	45995&	15028&	20085\\
\hline 
\textbf{Performance Ratio}&	5.66\%&	1.91\%&	3.50\%&	1.34\%&	2.10\%&	0.12\%&	0.56\%&	0.28\%\\
\hline

\end{tabular}
\vspace{0mm}
\vspace{-4mm}
\end{table*}

\section{Cryptographic Primitives}\label{crypto_primitive}
Trusted computing \cite{kauer2007oslo}, communication \cite{arif2018secure}, storage \cite{aujla2018secsva}, execution \cite{pinto2019demystifying}, and many other security objectives rely on cryptographic primitives including symmetric and asymmetric ciphers, data integrity algorithms, mutual trust key management and distribution mechanisms. As the application ecosystem evolves, new challenges and requirements of cryptographic primitives emerge. Figure \ref{fig_crypto_primitives} summarizes the cryptographic primitive implementation or performance optimization projects in the RISC-V domain. 

\subsection{Lightweight Cryptography}
First, small IoT devices require lightweight cryptographic algorithms. It is common that small IoT devices are CPU or memory resource-constrained, they may not be powerful enough to support efficient execution of standard cryptographic algorithms. 
As Table \ref{openssl_perf} shows, the latest 
Polarfire RISC-V FPGA has an average single-core compute performance of only 3.1\% of the Intel Xeon 4215 x86 processor.
Lightweight cryptography has been proposed to meet the requirements of resource constrained devices \cite{Campos2020}. Lightweight conceptually can refer to chip area size, code/memory size, energy efficiency, etc. There is an ISO/IEC 29192 standard for lightweight block ciphers. CLEFIA \cite{shirai2007128}, PRESENT \cite{lara2016novel}, LEA \cite{hong2013lea} are three of the ISO block cipher algorithms. NIST is also screening lightweight cryptography algorithms, which will be included into the NIST lightweight cryptographic standard. 
A major challenge of applying encryption in restricted environments is the trade-off between security and performance. Fabio et al. \cite{Campos2020} analyze different strategies for optimizing candidate solutions of the NIST lightweight cryptographic algorithms on the RISC-V architecture. Specifically, they demonstrated how multiple lightweight NIST candidates such as Gimli, Sparkle, Saturnin, Ascon, Delirium, and Xoodyak can be efficiently implemented. They studied the overall impact of optimizing symmetric key algorithms in assembly and C languages, proposing optimizations such as loop unrolling that can speed up the software-implemented algorithms by up to 81\%. 

\subsection{Symmetric and Asymmetric Cryptography}
Stoffelen et al. \cite{stoffelen2019efficient} highlight the features of RISC-V, and provide optimized assembly implementation of table-based AES, bit-sliced AES, ChaCha and Keccak-f[1600] for the RV32I instruction set. Regarding public key cryptography, they study the performance of arbitrary-precision integer arithmetic without the carry flag. They conduct quantitative performance studies on several RISC-V extensions, which provide design insights for future RISC-V core design and implementation.

Fixed function hardware accelerators such as an AES engine cannot support new ciphers. Kiningham et al. \cite{Kiningham2019Falcon} introduce Falcon, a hardware architecture used to accelerate various ciphers on energy-constrained devices. Falcon provides a general execution engine that support bitslice and permutation instructions, which are the backbone operations of current and probably future dominant ciphers including AES, Cha-Cha, SHA-256, RSA, Curve25519 ECC, and post-quantum cryptography such as R-LWE. For encryption technology, Falcon provides software flexibility while reducing the energy consumption of ciphers by 5 to 60$\times$ compared with software implementations. This improvement makes it feasible for IoT applications to upgrade the ciphers after deployment, so that they can always stay abreast of the latest security practices without shortening device deployment life or sacrificing application workload.

To deal with the security challenges of IoT devices, Loiseau et al. \cite{loiseau2018binary} propose a fast, low-power encryption technology for a new set of Binary Edwards Curves that have been defined to reach up to 284-bit security level suitable for IoT devices embedded with 32-bit general-purpose processors. They optimized the choice of point generators using \emph{w} coordinates to save multiplication in addition and doubling formulas. They managed to calculate one step of the Montgomery ladder with 4 multiplications and 4 squares. In addition to performance advantages, encryption on this 
curve also has inherent security properties against physical attacks.

\subsection{SIKE (Post-quantum Cryptography)}
Advances in quantum computers put the security of public key cryptosystems into risk \cite{chen2016report}. The security foundation of public key cryptosystems is that the integer-factorization problem is very complicated when the integer is very large. However, the quantum computer can solve this problem exponentially faster than traditional computers, which puts the existing public key cryptosystems at risk. Thus, if quantum computers are realized, public key cryptosystems may be easily compromised. In this background, NIST is screening public key encryption and digital signature algorithms that can resist potential attacks of quantum computers. 
The post-quantum algorithms will augment FIPS 186-4, SP 800-56A, SP 800-56B. 
NIST has selected 7 finalists and 8 alternates as candidates in their round 3 screening.
The algorithm evaluation will be finalized very soon.

SIKE public-key encryption and key-establishment algorithm is one of the NIST alternate candidates \cite{moody2020status}. It is a promising candidate standard, but its algorithms are resource-intensive. Although SIKE's FPGA implementation provides low latency and high performance, it has the disadvantages of large area and low flexibility. Compared with FPGA implementation, pure software implementation has much lower performance. 

Software and hardware co-design are important to optimize performance and satisfy design constraints such as cost and power consumption, as well as to shorten the time to market considerably \cite{teich2012hardware}. The openness of RISC-V provides unprecedented hardware optimization opportunities.
Roy et al. \cite{roy2020efficient} propose the hardware/software co-design method of SIKE, integrating the finite field accelerator based on redundant numbers into two microcontroller platforms based on Arm and RISC-V. The results show that, compared with the independent software implementation on Arm32 and Arm64, the implementation on the Arm Cortex-A9 enhanced with the field accelerator provides a significant speedup in terms of clock cycles. In addition, in order to show how to reduce the communication overhead between the processor and the accelerator, they directly integrate the finite field accelerator into the core of the RISC-V processor. This is the first hardware and software co-design method to implement SIKE design on Arm and RISC-V platforms. The proposed design requires 65500K clock cycles to execute SIKE on the Arm Cortex-A9 processor. On RISC-V, the proposed design only requires 36900K clock cycles.

\subsection{TRNG}
As we have discussed in Section \ref{RoT}, the random number generator is an important root of trust module. 
Saarinen et al. \cite{saarinen2020building} propose RISC-V true random number generator (TRNG) architecture that separates the entropy source component and the encrypted PRNG into a single interface. This is different from the previous TRNG implementations. They describe the interface and its use in cryptography, and discuss the background and basic principles of the interface. The design refers to the mainstream ISA, the latest SP 800-90B and FIPS 140-3 entropy review requirements, AIS-31 and Common Criteria for IT Security Evaluation, as well as current and emerging encryption requirements such as post-quantum encryption. The architecture choice is the result of quantitative observations on secure microcontrollers, Linux kernels and random number generators in cryptographic libraries. They further compare this architecture with some contemporary random number generators and describe a minimal TRNG reference implementation using an entropy source and RISC-V AES instructions.

\subsection{DICE}
The Device Identifier Composition Engine (DICE) is the minimal requirement for trusted computing on microcontrollers. Currently, most implementations use hardware that is not specifically designed for this purpose. These implementations rely on black-box MPUs. Because hardware that is not originally designed for DICE is used, there are certain pitfalls in the implementation process. Jager et al. \cite{Jager2020DICE} propose a DICE architecture based on a microcontroller, which is equipped with hardware that meets DICE requirements. It includes minor modifications to the processor pipeline, dedicated memory blocks, and modified interrupt and debug modules. They create an FPGA prototype based on the VexRiscV platform and evaluate the impact of the increase in chip size and DICE extension on the runtime to prove that DICE can be implemented with minimal changes to the microcontroller design and used in IoT and automotive environments as a trusted component.













%
\begin{figure}[t]%
    \centering
    {\includesvg[width=14cm]{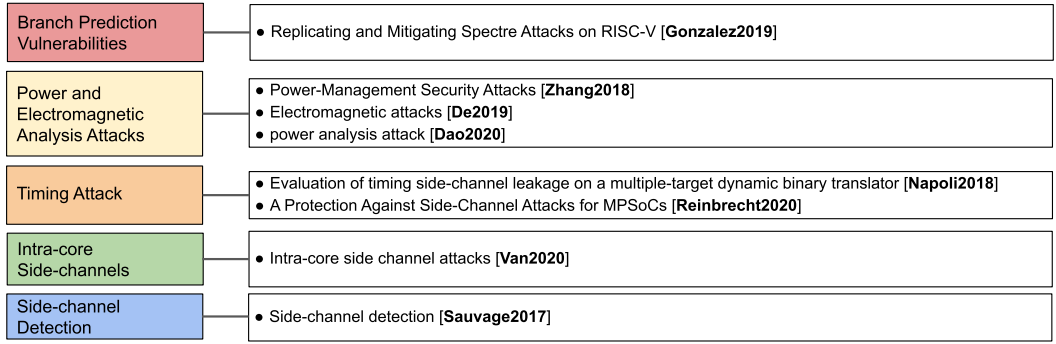}}
    \caption{Protection against Side-channel Attacks. \emph{References: \textbf{Deng2019}\cite{deng2019secure}, \textbf{Gonzalez}\cite{gonzalez2019replicating}, \textbf{Napoli2018}\cite{Napoli2018}, \textbf{Reinbrecht}\cite{Reinbrecht2020Guard}, \textbf{De2019}\cite{de2019protecting}, \textbf{Zhang2018}\cite{zhang2018blacklist}, \textbf{Dao2020}\cite{Dao2020}, \textbf{Van2020}\cite{van2020protecting}, \textbf{Sauvage2017}\cite{sauvage2017secure}}.}%
\vspace{-4mm}
    \label{fig_side_channel_prevention}%
\end{figure}

\section{Protection against Side-channel Attacks}\label{side_channel}
Side-channel attacks can be carried out through power analysis \cite{zhao2018fpga} and electromagnetic analysis (EMA) \cite{agrawal2002side}. In Section \ref{isa_side_channel_prevent}, we discussed the research on preventing side-channel attacks through ISA extensions. In addition to ISA extensions, there are many general methods such as implementing countermeasures in the CPU execution pipeline, or checking whether the hardware is in constant execution by analyzing the RTL code of the circuit. In this section, we discuss some general methods to detect and prevent side-channel attacks. Specifically, as Figure \ref{fig_side_channel_prevention} shows, we will discuss the branch prediction, TLB, inter-core, and timing side channels, and discuss the power analysis and electromagnetic attacks that exploit side channels. We will also discuss virtual prototyping, which can detect simple side channels.

\subsection{Branch Prediction Vulnerabilities}
The side-channel vulnerabilities of modern processors make hardware security a top priority in processor design. Gonzalez et al. \cite{gonzalez2019replicating} demonstrated how a general-purpose open source Berkeley Out-of-Order Machine (BOOM) based on a RISC-V processor can be used to research mitigating side-channel attacks at the microarchitecture level. First, they replicate several basic variants of the Spectre vulnerability \cite{kocher2019spectre}, which utilize speculative execution in the L1 data cache. Then, they implement preliminary hardware mitigation measures against this type of attack, prove its effectiveness, and measure its impact on performance and area size. Compared with the baseline processor, evaluation of the hardware mitigation shows that in the 45nm process, the IPC has increased by 2\%, the area has increased by 2.5\%, and the clock has been reduced by 0.36\%. This work confirms the value of the open source RISC-V hardware ecosystem for secure hardware research.

\subsection{Power and Electromagnetic Analysis Attacks}
Running unprotected software on an unprotected microprocessor can cause various side-channel leakages including direct-value leakage, data-overwrite, and circuit-level leakage \cite{de2019protecting}. The software implementation of cryptographic algorithms is vulnerable to side-channel analysis (SCA) attacks because the cryptographic key may leak through the processor's measurable physical characteristics such as power consumption and electromagnetic radiation. Various algorithms with solutions have been proposed to mitigate this issue. However, they rely on equipment assumptions that are rarely met, thus are not easy to implement. Moreover, these solutions do not consider microarchitecture-related issues. Mulder et al. \cite{de2019protecting} propose integrating the countermeasures of side-channel analysis into the RISC-V implementation. They use masking technology to wrap the secret before writing it to memory and do a reverse operation to unwrap the secret after reading it back from memory, so as to protect memory access from SCA. The solution prevents first-order power or electromagnetic attacks while keeping the implementation cost as low as possible. The evaluation results confirm the security of various encryption primitives running on the protected hardware platform.

Dynamic Frequency Scaling (DFS) is a technology related to the dynamic change of the clock frequency and usually the associated voltage of a CPU or hardware module during operation to adapt its power consumption \cite{bao2016static}. Dao et al. \cite{Dao2020} demonstrates how to integrate DFS technology into an open source RISC-V processor and use it as a simple, cost-effective countermeasure against Simple Power Analysis attacks. The idea is that the DFS module can conceal sensitive information in the measured power trace, while the processor's hardware resource requirements hardly change. They implemented DFS in the Sakura-X FPGA board and demonstrated the usefulness of this method to mitigate Simple Power Analysis attacks. 

Most modern computing devices can control frequency and voltage through fine-grained operations. CLKSCREW \cite{tang2017clkscrew} is a new type of fault attack, which uses the security obliviousness of the energy management mechanism to make the frequency of the equipment exceed its operating limit, which leads to malfunction and security violation.
Statically and permanently limiting the frequency and voltage modulation space can mitigate such attacks, but will result in a significant drop in performance and energy efficiency. Zhang et al. \cite{zhang2018blacklist} propose a runtime technology that uses a neural network model to dynamically blacklist unsafe operating performance points. The model is first trained offline at design time, and then adjusted at runtime by checking a set of selected functions such as power management control registers, timing error signals, and core temperature. They designed an algorithm and hardware called the BlackList (BL) core, which can detect and mitigate this kind of power management-based security attacks with high accuracy. The BL core generates a relatively small amount of overhead in terms of power, delay, and area size.
%
\subsection{Timing Attack}
The calculation of algorithms takes time. If the algorithm is not carefully designed with constant execution time, it may cause different inputs to have different execution times. If the calculation involves sensitive information such as a cryptographic key, the attacker may reverse the content of the key by traversing a large number of input vectors and accurately measuring the execution time of each test vector, causing the leakage of sensitive information. Prying sensitive key data through time information is usually much easier than cryptanalysis using known pairs of plaintext and ciphertext. 

The Translation Lookaside Buffers (TLBs) in modern processors may cause timing side-channel attacks. But exploiting the TLB channel is challenging due to unknown addressing functions inside the TLB and the
attacker's limited monitoring capabilities that at best cover only the victim's coarse-grained data accesses. However, recently researchers can reverse engineer the addressing functions inside the TLB, and devise a machine learning strategy that exploits high-resolution temporal features about a victim's memory activity, which make TLB side-channel attack practical \cite{gras2018translation}. To mitigate this risk, Deng et al. \cite{deng2019secure} introduce a novel three-step modeling method that is used to exhaustively enumerate all possible TLB-based timing vulnerabilities. Step 1 performs memory operations and places the TLB block in a known initial state. Then, step 2 performs the second memory operation to change the state of the TLB block. Finally, Step 3 performs the final memory operation, and the time of the final operation depends on the addresses of Step 1, Step 2 and Step 3. Attacks with more than three steps can be broken down into a three-step attack.
Based on the three-step model, they show how to automatically generate a micro-security benchmark that can test TLB vulnerabilities. They propose two new secure TLB designs: static partition (SP) TLB and random fill (RF) TLB. The evaluation of the secure TLB implemented in RISC-V Rocket Core shows that the new TLB can not only defend against previously announced attacks, but also against other new timing-based attacks in the TLB discovered using the new three-step model. The FPGA-based evaluation shows that the RF TLB defends against all attacks with performance overhead of less than 10\%.

Multi-processor system-on-chip (MPSoC) is a popular computing platform suitable for various applications due to its energy efficiency and flexibility. SoCs with heterogeneous architecture allow for the integration of various central processing units and even graphics processors on the same system are getting popular \cite{ruaro2019memphis}.
Like many other platforms, they are also vulnerable to side channel attacks (SCA). Logical SCA can retrieve sensitive information by simply observing the system attributes that depend on the software executed by the victim on the MPSoC, which is very harmful. Unfortunately, many current protection mechanisms are either platform-dependent or only effective against a few attacks. Reinbrecht \cite{Reinbrecht2020Guard} introduces Guard-NoC, which is a secure network-on-chip (NoC) architecture that can protect MPSoC from various Logical SCAs. The secure NoC uses three application-independent strategies to hide and isolate sensitive information by masking the execution time of the operation and employing dual communication strategy such as using packet and circuit switching at the same time. Packet switching is used for secure packets and circuit switching is used for common packets. Evaluation shows that the security NoC can resist the actual Logical SCAs, and hardly leak any information, while having a minimal area size and power consumption.

Timing side channel attacks are an important issue in cryptographic algorithms. If the execution time of the implementation depends on secret information, the adversary can infer the secret by measuring the execution time. Different methods have recently emerged to explore information leaks in encryption implementations and protect them from these attacks. For example, in Section \ref{subsect_hardware_physical_security} we have discussed IODINE \cite{gleissenthall2019iodine}, which translates Verilog code for a formal analysis to detect timing channels. However, there is very little about ISA emulation and its impact on timing attacks. Napoli et al. \cite{Napoli2018} studied the impact of OI-DBT, a dynamic binary translator using different region formation technologies (RFT), on the implementation of constant time and non-constant time of cryptographic algorithms. Experiments show that simulation can have a significant impact on secret leakage, and even mitigate leakage in some cases. In addition, the results show that the simulator's choice of RFT heuristic also has an impact on these leakages.

In Section \ref{hardware_assist_sec_unit}, we have discussed program obfuscation as a widely used intellectual property (IP) protection technique against reverse engineering attacks. 
Biswas et al. \cite{biswas2021cryptographic} observed that the choice of transformation sequence has a significant impact on  timing channel information leakage. Certain transformation sequences may cause leakage higher than the original program. Biswas et al. proposed a timing channel sensitive program obfuscation optimization framework based on the genetic algorithm to find the best combination of obfuscation transformation functions in terms of performance and prevention of timing side channel leakage. They evaluated the new framework on the RISC-V Rocket core. They use the \emph{dudect} tool to verify that the proposed TSC-SPOOF framework provides optimization points for ModExp and MulMod16 programs, while reducing timing channel leakage. They observed that for the optimized ModExp and MulMod16 programs, it takes about 1M and 2M measurements, respectively, to pass through the \emph{t}-statistic of 10. But for the two programs in the initial population, only 20K measurements are needed to cross the same \emph{t}-statistic. 


\subsection{Intra-core Side-channels}
Systems that protect the enclaves from privileged software must consider software-based side-channel attacks. Protection against attacks that can be launched from privileged software is an emerging attack model. Van et al. \cite{van2020protecting} propose to protect against intra-core side-channel attacks by enforcing that all active enclaves are physically isolated on their own separate core, thereby mitigating side channel attacks inside the core. They also redesign the memory hierarchy based on the ownership of the security zone to protect the security zone from intra-core side channel attacks. The combination of physical isolation and a redesigned memory hierarchy can protect the enclave from all known software-based side-channel attacks.
The memory tag is used to protect the confidentiality and integrity of the enclave's memory. Bootstrapping Shim is the RoT, from which the management shim can start to manage the tags that protect the memory. The management shim is the software part of the TCB. The hardware forces enclaves to only access the pages they are authorized to access, and the management shim is the only code that is allowed to change the value in the tag directory. The management shim is not an operating system, it only implements the minimum logic required to securely maintain the enclave life cycle and transfer page ownership.
They implemented the system and evaluated it with communication performance, memory overhead, and hardware area metrics. Evaluation shows that adding secure cores to a modern system on chip would increase CPU complexity by about 14.5\%, and increase the hardware area by less than 2\%. Storing the extra tag data increases the size of the last-level cache by 2\% to 13\%.

\subsection{Side-channel Detection with Virtual Prototyping}
Evaluating software security vulnerabilities in the design phase can find problems at the earliest stage and avoid the cost of later vulnerabilities repair, which is therefore very critical. Sauvage et al. \cite{sauvage2017secure} describe a virtual prototype implemented by scalar multiplication, aiming to protect a platform from simple side-channel attacks. They used the Mentor Graphics Modelsim tool to obtain a reproduction of information leakage as close to reality as possible, requiring bit and clock cycle accuracy to simulate the execution of the software implementation on PULPino \cite{traber2016pulpino}, which is an open source 32-bit RISC-V microcomputer. For each clock cycle, they calculate the number of bits entering the microcontroller, the power consumption image, and observe the program counter to identify the executed assembly instructions, and then identify the corresponding C function. They use naive double-and-add implementation relying on cryptographic primitives of the mbed TLS library for reference. Virtual analysis points out that there are differences between the \emph{double} function on one side and the \emph{add} function on the other side in the way of managing variables and internal operations, which can be exploited to extract the private key. This method is still immature, and it faces many challenges to get practical applications, such as improving simulation performance, more realistic attack models, and more automated deployment.


\section{Conclusion and Future Work}\label{future_directions}
Hardware-based security technologies such as memory protection and instruction set security extensions have been widely used in practice. Since the Meltdown and Spectre vulnerabilities broke out, the security of computer hardware and architecture has received extensive attention from industry and academia, attracting more and more research. As a new open instruction set, RISC-V has received extensive attention, and is moving towards the mainstream. Although there are more and more research projects related to RISC-V, the security research of RISC-V as a new architecture is relatively lagging. In this article, we investigate the current status of RISC-V hardware and architecture security research, hoping to provide readers with a big picture of the RISC-V security ecosystem. Specifically, we first briefly describe the background and architectural security foundation of RISC-V, and then introduce the security research of RISC-V by topic. We focus on discussing hardware and architecture security. Our research covers hardware and physical access security, hardware-assisted security units, ISA security extensions, memory protection, cryptographic primitives, and side-channel attack protection. During the investigation, we notice that the hardware and architecture foundation of RISC-V security is in place, and a wide range of security mechanisms have been established.

We also notice that the cryptographic instruction set of RISC-V is not complete, and the standardisation work is still in progress. The instruction set is the biggest space for improvement of RISC-V in the architecture layer. Realizing efficient cryptographic algorithms through the instruction set will be significant work. In addition, the security foundation of RISC-V architecture, such as PMP, may be attacked via side-channels \cite{nashimotobypassing2020}. Preventing the architectural security foundation from being attacked by the side-channels will be an important research topic. As we mentioned above, RISC-V is a new architecture, and some research is relatively lagging behind. Porting classic security applications of other architectures to RISC-V will be major and important tasks in the near future. For example, the RISC-V community is porting Arm's OP-TEE implementation. It will be meaningful to compare RISC-V with other architectures such as Arm in related fields. In addition, we can also see from our summary that although the current RISC-V research has covered a wide range of security topics, many researches are still in an early stage and related implementations are not yet mature. For example, the current research on memory protection is mainly focused on tagged memory, only a few researches are on memory isolation and encryption verification. Research on logic locking, electromagnetic injection attacks, side-channel prevention and detection, and control flow integrity is still immature. RISC-V security topics still have broad research space. In this article, we mainly discuss the hardware and architecture security of RISC-V.
As the next step, we will continue to conduct a comprehensive study of RISC-V firmware and system, as well as software and application security, in order to achieve complete coverage of RISC-V security spectrum.

As far as we know, this is the first comprehensive review article on RISC-V security. We hope that this article will give readers a comprehensive understanding of existing RICS-V security mechanisms, and even from a broader perspective, understand the state-of-the-art research on embedded systems and general-purpose computer architecture security. 



\printbibliography
\end{document}